\documentclass[prd,showpacs,showkeys,preprintnumbers,floatfix,
nofootinbib,superscriptaddress]{revtex4-1}
\usepackage{float}
\usepackage[utf8]{inputenc}
\usepackage{nicefrac}
\usepackage{mathtools}
\usepackage{amsfonts} 
\usepackage{amssymb} 
\usepackage{amsmath} 
\usepackage{graphicx} 
\usepackage{subfigure} 
\usepackage{array} 
\usepackage{dcolumn} 
\usepackage{bm} 
\usepackage{latexsym} 
\usepackage{longtable} 
\usepackage{hyperref} 
\usepackage{verbatim}
\usepackage{epsfig}
\usepackage{color}
\newcommand{\DL}{\mathcal D_L}
\newcommand{\LL}[0]{{\big (} \ \ 
\, \mathcal L_L^{(u)} \ \,  \ 
{\color{black}{\big )}}}
\newcommand{\RL}[0]{\Bigg( \! \hspace{1pt}  \mathcal R_L^{(u)} \!  \Bigg)}

\newcommand{\AProw}[0]{{\big (} 
\ \ \,  A'^{[B_2]} \ \,  \ {\big )}
}
\newcommand{\Acol}[0]{\Bigg ( \! \hspace{1pt} A^{[B_2]} \!  \Bigg)}

\newcommand{\APrhorow}[0]{{\big (} 
\ \ \, A'^{[B_2,\rho]} \ \,  \ 
{\big )}    
}
\newcommand{\Arhocol}[0]{\Bigg ( \! \hspace{1pt} A^{[B_2,\rho]} \!  \Bigg)}

\newcommand{\FthM}[0]{\Bigg ( \ \  i F_3 \ \  \Bigg )}

\newcommand{\rhoM}[0]{\Bigg ( \ \   \frac{i\rho}{2 \omega} \ \  \Bigg )}
\newcommand{\rhoMT}[0]{\Bigg ( \ \   \frac{i\rho}{2 \omega} \ \  \Bigg )^{\!\!\! \mathrm{T}} \ }

\newcommand{\Kdf}[0]{\mathcal K_{\mathrm{df},3}}

\newcommand{\Mth}[0]{\mathcal M_{ 3}}
\newcommand{\ML}[0]{\mathcal M_{ 2,L}}
\newcommand{\MthL}[0]{\mathcal M_{ 3,L}}
\newcommand{\K}[0]{\mathcal K_2}
\newcommand{\M}[0]{\mathcal M_2}

\newcommand{\PV}[0]{\widetilde{\mathrm{PV}}}
\begin{document}
\title{Expressing the three-particle finite-volume spectrum
in terms of the three-to-three scattering amplitude}
\author{Maxwell T. Hansen}
\email[e-mail: ]{hansen@kph.uni-mainz.de}
\affiliation{
Institut f\"ur Kernphysik and Helmholz Institute Mainz, Johannes Gutenberg-Universit\"at Mainz,
55099 Mainz, Germany\\
}
\author{Stephen R. Sharpe}
\email[e-mail: ]{srsharpe@uw.edu}
\affiliation{
 Physics Department, University of Washington, 
 Seattle, WA 98195-1560, USA \\
}
\date{\today}
\begin{abstract}
In this article we complete our formalism relating the finite-volume
energy spectrum of a scalar quantum field theory to the three-to-three
scattering amplitude, $\Mth$. In previous work (Ref.~\cite{us}) we found a
quantization condition relating the spectrum to a non-standard
infinite-volume quantity, denoted $\Kdf$. Here we present the relation
between $\Kdf$ and $\Mth$. We then discuss briefly how our now completed
formalism can be practically implemented to extract $\Mth$ from the
finite-volume energy spectrum.

\end{abstract}
\maketitle

\section{Introduction}

This work completes the formalism that was partially developed in
Ref.~\cite{us}. In that work we determined the finite-volume spectrum
of a relativistic scalar quantum field theory by 
calculating an appropriate finite-volume correlator to all orders in
perturbation theory. All contributing Feynman diagrams were
decomposed into products of
finite- and infinite-volume quantities. Summing these factored
expressions gave a result for the correlator in terms of
finite-volume kinematic functions and infinite-volume scattering
quantities. The poles in the resulting expression determine the
energies of the finite-volume states, and identifying their locations
gave a relation between the finite-volume spectrum and infinite-volume
scattering quantities.

The central drawback of the result of Ref.~\cite{us} is that it
depends on a nonstandard infinite-volume three-particle quantity, a
modified three-particle K-matrix denoted $\Kdf$. 
For our formalism to be useful, it is necessary to
relate $\Kdf$ to the
physical three-particle scattering amplitude $\Mth$.
This is what we achieve in the present work.
This shows that, in a relativistic context, the finite-volume
spectrum of three particles is determined by infinite-volume observables
(up to corrections falling faster than any power of $1/L$,  where $L$ is the linear extent of the finite volume).
This result was previously established using nonrelativistic
effective field theory~\cite{Polejaeva:2012ut}.

\begin{figure}[bh]
\begin{center}
\includegraphics[scale=1.2]{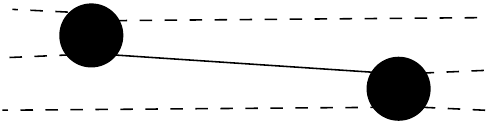}
\caption{Example of a singular contribution to the on-shell
  three-to-three scattering amplitude.  Dashed lines are on-shell,
  amputated, external propagators, while the solid line is a fully
  dressed propagator. Filled  circles represent two-to-two scattering 
  amplitudes.  The momentum flowing along the internal
  (solid) line goes on shell (leading to a divergent scattering amplitude)
  for particular choices of external momenta
  corresponding to two isolated two-to-two scattering events.}
\label{fig:singdiagram}
\end{center}
\end{figure}

To give a sense of what is involved in relating $\Kdf$ to $\Mth$,
we recall from Ref.~\cite{us} the major differences between the
two quantities. Both are obtained by summing contributions
from diagrams with six external lines.
However, divergences that are present in $\Mth$ are removed from $\Kdf$
(leading to the subscript ``df'' for ``divergence free''). 
These divergences are due to diagrams containing
pairwise scatterings, such as that shown in Fig.~\ref{fig:singdiagram}. They have nothing to do with bound
states but are instead a result of on-shell three-particle intermediate 
states~\cite{Taylor:1977A,Taylor:1977B,Brayshaw:1969ab,Rubin:1966zz}. 
It is not surprising that the finite-volume
spectrum is related to a modified quantity in which these divergences are
removed. We stress that the terms removed depend only on the on-shell
two-to-two scattering amplitude, $\M$. These can thus be separately
computed and added to the divergence-free quantity to recover $\Mth$.

\begin{figure}[tbh]
\begin{center}
\includegraphics[scale=0.3]{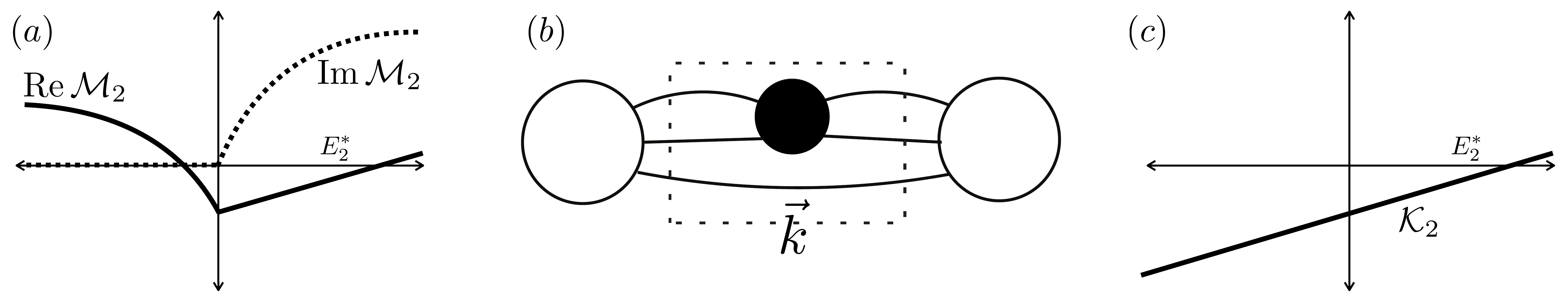}
\caption{(a) Schematic plot of the the s-wave two-to-two scattering
  amplitude, $\mathcal M_{2,s}$, as a function of two-particle CM
  energy, $E^*_2$, showing the well known cusps at threshold 
  ($E^*_2=2m$) in both real and imaginary parts.
   (b) Example of two-to-two
  scattering as a subprocess of three-to-three scattering. Because
  total energy and momentum are fixed, the energy and momentum of the
  top two propagators varies with the momentum of the bottom
  propagator. Thus the sum over $\vec k$ has important finite-volume
  effects arising from the two-particle unitary cusp. 
  (c) In contrast to $\M$, $\K$ is a smooth function at threshold. 
  Working with $\K$ in place of $\M$ resolves the issue of 
  finite-volume effects from threshold cusps.}
\label{fig:cusps}
\end{center}
\end{figure}

A second and more important
difference between $\Mth$ and $\Kdf$ concerns the
pole prescription used to define momentum integrals over products of
propagators. In $\Mth$ the standard $i\epsilon$ prescription is used,
while $\Kdf$ uses
a modified principal value prescription (denoted $\PV$).
The use of a nonstandard pole prescription is required
in order to remove the unitary threshold cusp in the
two-particle scattering amplitude, 
as was pointed out in Ref.~\cite{Polejaeva:2012ut}. 
As described in detail in Ref.~\cite{us} and sketched in Fig.~\ref{fig:cusps},
such cusps, if not removed, generate important finite-volume effects
when two-to-two scattering is considered as a subprocess of
three-to-three scattering.
Our prescription avoids such cusps, but at the cost of
introducing a nonstandard infinite-volume quantity.

 A third difference between $\Mth$ and $\Kdf$ is that the latter is 
defined using an ultraviolet cut-off (the function $H(\vec k)$ defined in Eqs.~(\ref{eq:HvalA})-(\ref{eq:Jdef}) of the appendix)
while $\Mth$, being a physical quantity, is independent of the cut-off function.

The final difference is that $\Kdf$ 
involves contributions in addition to the standard Feynman diagrams,
which we denote ``decorations''. These make the definition of $\Kdf$
very complicated, as explained in Ref.~\cite{us}.

In this work, we relate these two quantities using an indirect approach.
By extending the methods of Ref.~\cite{us},
we obtain a relation between $\MthL$, which is a finite-volume version of $\Mth$, and $\Kdf$, Eq.~(\ref{eq:MthLres}) below.
The desired relation  between $\Mth$ and $\Kdf$ is then reached by taking the infinite-volume limit.
This approach avoids the need to directly reference the
very complicated definition of $\Kdf$.
As a side benefit, we obtain some intuition about the properties of $\Kdf$ itself.

Combining the results of Ref.~\cite{us} with the present article gives a complete relation between the finite-volume energy spectrum and two-to-two and three-to-three scattering amplitudes. The result relies on no assumptions concerning the details of the underlying theory, no specific effective field theory and no power-counting scheme. Further, the relation is derived in fully relativistic field theory, and thus does not rely on a small momentum expansion.  However, as it stands, the formalism is limited in two important ways.
First, it can only describe systems of identical scalar particles with a $\mathbb Z_2$ symmetry that prevents the
coupling of states containing even and odd numbers of particles. In particular, no two-to-three scattering
is included.
Second, there can be no two-particle resonances in the energy range considered. If the center of mass (CM) energy of the three-particle system is $E^*$, then resonances must not appear
below two-particle CM energies of $E^*-m$.\footnote{%
Since $E^*$ is the CM energy for three particles, $E^*-m$ is the maximum possible energy for a two-particle subsystem.}
We make this restriction more precise in the paragraph following Eq.~(\ref{eq:qkstardef}) below. 
There is an additional, less restrictive condition, namely that $E^*$ must 
lie above the single-particle pole and below the five-particle production threshold ($ m < E^* < 5m$). 
This is analogous to the restriction of the two-particle quantization condition to energies below
the inelastic threshold.

Despite these restrictions, there are several physical systems to which our formalism can be directly applied.
In the context of QCD, one must consider three-meson systems with a quantum number that enforces the
$\mathbb Z_2$ symmetry.
For example, in the limit that isospin is an exact symmetry, states with even and odd numbers of pions
decouple due to G parity. In this case one can project onto states of definite $I$ and $I_3$ and
effectively obtain a system of three identical scalars. A simple example is the $\pi^+\pi^+\pi^+$ system
with $I=I_3=3$. 
For all choices of isospin, our formalism is directly applicable up to the five pion threshold. The resonances that are present in some channels,
e.g. the $\rho$ if there is an $I=1$ two-particle subchannel, occur at energies above the five pion threshold.\footnote{We caution that for sufficiently heavy pions the $\rho$ resonance is pushed below the four particle threshold of the two-to-two system, and thus below the five-particle threshold in the three-to-three system. In this case the rho mass, rather than the $5m$ threshold, becomes the upper kinematic bound.}

Another application of our present formalism is to study three kaon and three $D$ meson systems, for which the conservation of strangeness and charm in QCD prevents coupling to other states. 

Finally, we note that our formalism 
also applies for constituent particles such as ultra-cold atoms and molecules.
Here a non-relativistic description is usually expected to work very well so that 
the relativistic machinery we employ would appear to be overkill. However, in certain cases the 
relativistic result can give interesting information even very close to threshold. 
For example in a recent work, Ref.~\cite{ourpt}, we have studied the threshold 
shift of two- and three-particle states in $\lambda \phi^4$ theory. We expand 
the shift from threshold in powers of inverse box length, i.e.~powers of $1/L$, and 
find that at $1/L^6$ the result differs from previous non-relativistic analysis~\cite{Beane:2007qr}. 
This is also the order at which three-body couplings first arise in the expansion. 
Thus, in this context, three-particle and relativistic effects have the same volume 
scaling and it is natural to include them together. 

Going beyond perturbative expansions, we also note that our formalism allows for arbitrarily large two-particle scattering lengths. In particular, our results can be used to describe systems at unitarity, where the two-particle scattering length diverges.\footnote{The scattering length $a$ and effective range $r$ are defined via
\begin{equation*}
p \cot \delta(p) = - \frac{1}{a} + \frac{1}{2} r p^2 + \mathcal O(p^4) \,,
\end{equation*}
where $\delta$ is the s-wave scattering phase shift and $p$ is the magnitude of three momentum for one particle in the two-particle CM frame. Our formalism is only valid if $p \cot \delta(p)$ does not pass through zero for a particular $p$-value in the energy range considered. In the context of the unitary limit, this requires setting $r$ and all higher order coefficients to zero before sending $a \rightarrow \infty$.} In this vein an interesting study has recently been published in Ref.~\cite{Meissner:2014dea}. Here the authors consider a non-relativistic three-particle system with two-particle scattering at unitarity. They then analyze how a shallow three-particle bound state would be shifted due to finite-volume effects. Their result is a clear quantitative prediction for the leading finite-volume shift, and reproducing this with the present general formalism would be an interesting and important check.

The remainder of this article is organized as follows.
In the following section, we recall the results of Ref.~\cite{us} that are needed here.
In Secs.~\ref{sec:B2} and \ref{sec:B3} we derive the relationship between
$\MthL$ and $\Kdf$. The former section includes only two-particle Bethe-Salpeter
kernels, while the latter adds in three-particle kernels.
In Sec.~\ref{sec:limit} we take an appropriate infinite-volume limit of $\MthL$ and
obtain the desired relation between $\Mth$ and $\Kdf$.
We then show, in Sec.~\ref{sec:KfromM}, how to invert this relationship so that
$\Kdf$ can be obtained from $\Mth$.
Section~\ref{sec:simp}  shows how the general relation between these quantities simplifies in two approximations
studied in Ref.~\cite{us}. We conclude in Sec.~\ref{sec:conc}.
We include an appendix which collects the definitions of key quantities from Ref.~\cite{us}.

\section{Finite-volume correlator from previous work}
\label{sec:corr}

In this section we summarize the relevant results from Ref.~\cite{us},
and rewrite them in a form useful for our subsequent analysis.
We also introduce the finite-volume three-to-three scattering
amplitude $\MthL$, the quantity that plays a central role in our analysis.

We begin by reviewing the set-up for the calculation presented in
Ref.~\cite{us}. We work with a quantum field theory containing a
real scalar, $\phi$, having physical mass $m$. The theory is assumed
to have a $\mathbb Z_2$ symmetry so that only even powers of $\phi$
appear in the Lagrangian and thus even- and odd-particle states have
zero overlap. The derivation of Ref.~\cite{us}
also requires that the total center-of-mass (CM) energy is such
that only three-particle states can go on-shell, and also such that there are no resonances in the two particle subchannels of the three-particle states. 
The theory is otherwise assumed to be completely
general. In particular, it need not be renormalizable and can
contain all possible even-legged vertices with no assumptions about
relative coupling strengths.

This theory is then considered in a finite, cubic spatial volume of
extent $L$ with periodic boundary conditions. The time extent is
taken infinite.  
In this geometry, only spatial momenta whose components are 
integer multiples of $2 \pi/L$ can propagate. 
The volume extent $L$ is also assumed to be
large enough that exponentially suppressed corrections of the form
$e^{-mL}$ can be neglected. Such terms are dropped throughout the
derivation. Power-law corrections proportional to powers of
$1/(mL)$ are kept to all orders.

The analysis of Ref.~\cite{us} considers a finite-volume correlator defined by
\begin{equation}
\label{eq:corrdef}
C_L(E, \vec P) \equiv \int_{L} d^4 x\; e^{i(E x^0-\vec P \cdot \vec x)}
\langle 0 \vert \mathrm{T} \sigma(x) \sigma^\dagger(0) \vert 0 \rangle
\,.
\end{equation}
Here T indicates time-ordering while \(\sigma(x)\) and
\(\sigma^\dagger(x)\) are interpolating fields coupling to states with
an odd number of particles. The Fourier transform restricts the intermediate states contributing to the correlator to have
total energy $E$ and momentum $\vec P\in (2 \pi/L)\mathbb Z^3$. 
Thus the energy in the center of mass (CM) frame is
$E^*\equiv\sqrt{E^2-\vec P^2}$.
Our analysis concerns the region $m<E^*<5m$,
where only three-particle states contribute.
Note that, although we work in Minkowski space, 
the finite-volume spectrum is not
affected by this choice. Our result is applicable even if the energies
are determined in some other way, for example by using a Euclidean
correlator as in a lattice QCD calculation.

\begin{figure}
\begin{center}
\includegraphics[scale=0.42]{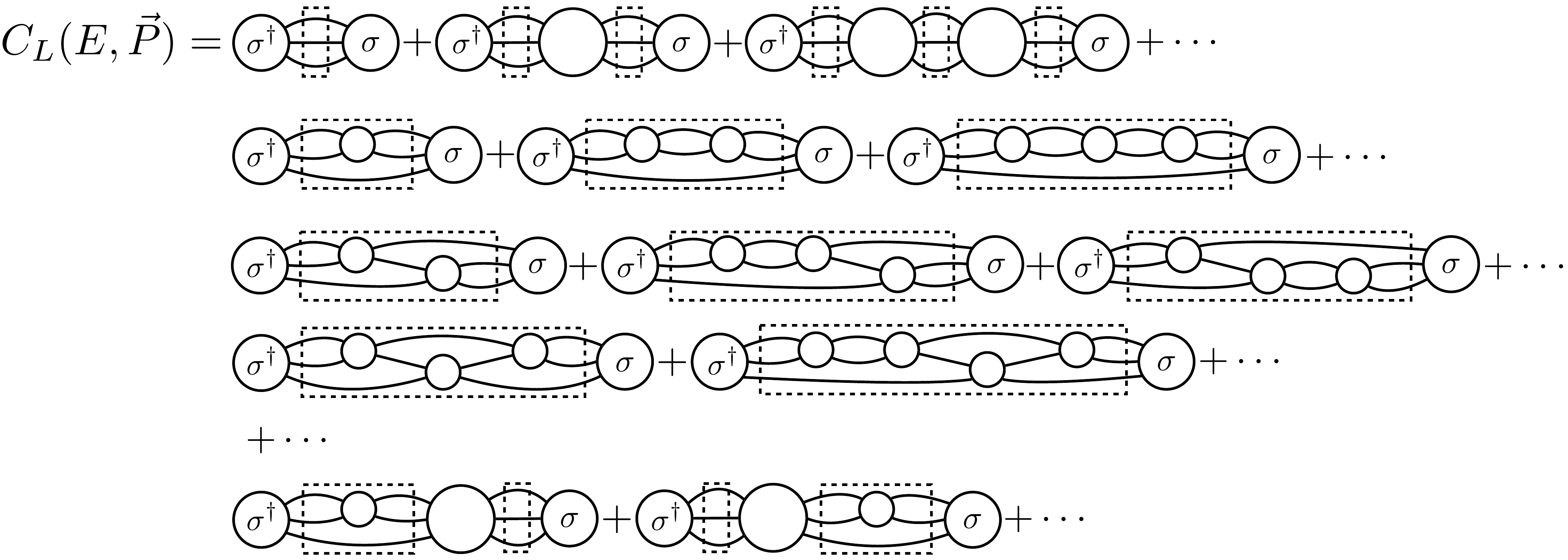}
\caption{Skeleton expansion for the finite-volume correlator. 
  Outermost blobs in all diagrams represent functions of momentum 
  that are determined by the interpolating operators 
  $\sigma$ and $\sigma^\dagger$. 
  Insertions between these functions having four legs represent two-to-two
  Bethe-Salpeter kernels, \(i B_{2}\), while insertions with six legs
  represent the analogous three-to-three kernels, \(i B_{3}\). 
  Lines connecting kernels and \(\sigma\)-functions represent
  fully-dressed propagators. As explained in the text, the kernels and
  dressed propagators can be replaced by their infinite-volume 
  counterparts (in which internal loop momenta are integrated).
  However, the spatial momenta flowing along the propagators that are
  shown explicitly, and which lie
  within the dashed rectangles, are summed rather than integrated.} 
\label{fig:fullskelexpansion}
\end{center}
\end{figure}

The finite-volume correlator, $C_L$, has the skeleton
expansion shown in Fig.~\ref{fig:fullskelexpansion}. 
This is explained in Ref.~\cite{us} (where precise definitions
of the kernels $B_2$ and $B_3$ can also be found).
Here we only note the essential features.
Note first that $C_L$ differs from its infinite-volume counterpart, $C_\infty$,
only in that spatial loop momenta are summed rather than integrated.
The key observation is then that
replacing sums with integrals leads to power-law finite-volume corrections
only when all particles in a cut can go on shell.
Since we have chosen $m<E^*<5m$, this can occur only for three-particle cuts.
For loops with cuts involving five or more particles one makes only
exponentially suppressed errors when replacing sums with integrals.
As we neglect such corrections throughout,
the Bethe-Salpeter kernels (for which cuts contain either one or else five or more particles)
can be replaced by their infinite-volume counterparts.
The same holds true for the dressing functions of the propagators.
Thus the only impact of working in finite volume is that the propagators
contained within the dashed rectangles in the figure have
spatial momenta that are summed.

The task accomplished in Ref.~\cite{us} is to factorize the power-law
finite-volume dependence contained in $C_L$ from quantities defined 
in infinite volume. The end result is
\begin{equation}
\label{eq:corrresult}
C_L(E, \vec P) = C_\infty(E, \vec P) + i A' \frac{1}{1+ F_{3} \Kdf}
F_{3} A \,.
\end{equation}
Here $A'$, $A$ and $\Kdf$ 
(as well as $C_\infty$) are infinite-volume quantities,
while $F_3$ depends on $L$ [as well as on $E$, $\vec P$ and the
two-particle scattering amplitude---see below].
The result (\ref{eq:corrresult})
is a matrix equation, in which  $A'$ is a row vector, 
$\Kdf$ and $F_3$ are matrices and $A$ is a column vector.
The index space is a direct product of 
two-particle angular momentum, parametrized by
spherical harmonic indices $\ell$ and $m$, together with a discrete
finite-volume momentum, $\vec k \in (2 \pi/L)\mathbb Z^3$. 
So, for example, 
$\Kdf = \mathcal K_{\mathrm{df},3; k' \ell' m'; k \ell m}$ 
where $k'$ and $k$ are shorthand for discretized three-vectors.

As we will see below, $C_\infty$, $A$ and $A'$ do not appear in the
main results of this paper and so we do not recapitulate 
their definitions from Ref.~\cite{us}.
As for $\Kdf$, the main task of this paper, as noted in the introduction,
is to come up with an alternative to the very complicated and implicit
definition given in Ref.~\cite{us}.
Thus at this stage we only present the definition of $F_3$, which is
\begin{equation}
\label{eq:FthdefML}
i F_3 \equiv \frac{i F}{2 \omega L^3} \left[\frac13 + \frac{1}{1 - i
    \ML i G} i \ML i F\right ] \,.
\end{equation}
Like $F_3$, the quantities $1/(2 \omega L^3)$, $F$, $G$, $\ML$ are all
matrices with indices $k' \ell' m';k \ell m$. 
The first three of these quantities are known kinematic functions that depend on $L$
(as well as $E$, $\vec P$, and $m$). 
They are defined in the appendix.
The final quantity is the finite-volume, two-to-two
scattering amplitude, $\ML$. This in turn can be expressed 
(up to exponentially suppressed corrections) as
\begin{equation}
\label{eq:M2toK2}
i \ML \equiv \frac{1}{1 - i \K i F} i \K \,,
\end{equation}
where $\K$ is a modified two-particle infinite-volume K-matrix introduced in 
Ref.~\cite{us}.
The relation (\ref{eq:M2toK2}) is needed to show the equivalence of the result
(\ref{eq:FthdefML}) to the expressions for $F_3$ given in Ref.~\cite{us}. 

Since $\ML$ was barely used in Ref.~\cite{us}, but plays an important
role here, we discuss its properties in some detail.
A useful, although imprecise, way of thinking about $\ML$ is as
the two-particle scattering amplitude {\em in finite volume}. 
Recall that $\ML$ has indices $k' \ell' m';k \ell m$,
with $k$ and $k'$ short for $\vec k$ and $\vec k'$, respectively.
For a two-particle quantity such as $\ML$, the momenta $\vec k$ and
$\vec k'$ are those of the third, spectator particle, which is 
unscattered. Thus $\vec k=\vec k'$, or, in matrix notation,
${\cal M}_{2,L;k' \ell' m';k \ell m} \propto \delta_{k' k}$.
The same holds for $\K$ and $F$.
Thus the spectator momentum serves only to determine the energy-momentum
flowing through the scattered pair, which is
\begin{equation}
\label{eq:P2def}
P_2 = (E-\omega_k,\vec P-\vec k)\,, \ \  \textrm{with}\ \ 
\omega_k= \sqrt{\vec k^2+m^2}\,.
\end{equation}
The energy of the two scattered particles in their CM frame is therefore
\begin{equation}
\label{eq:Eskstardef}
E_{2,k}^* \equiv \big [ P_2^2 \big ]^{1/2}
= \big [(E - \omega_k)^2 - (\vec P - \vec k)^2 \big]^{1/2} \,,
\end{equation}
while the magnitude of their momenta in this frame is
\begin{equation}
\label{eq:qkstardef}
q_k^* = \big [E_{2,k}^{*2}/4-m^2 \big]^{1/2}\,.
\end{equation}
The only unconstrained degrees of freedom are the incoming
and outgoing directions of one of the particles in their CM frame,
which we denote $\hat a^*$ and $\hat a'^*$, respectively. 

We can now make precise the requirement mentioned above, 
namely that the theory should not have any two-particle resonances. 
The condition is that $\K$ should not have poles in the range of allowed two particle energies.
This requirement is most stringent for $\vec k=0$, since this maximizes the energy in the two particle subsystem.
Thus $\K$ should have no poles in the range of two particle CM energies 
$0 \le E_{2,0}^*\le E^* - m$. 
Such poles would lead to power law finite-volume effects when converting sums to integrals
in diagrams like that shown in Fig.~\ref{fig:cusps}(b), and such effects are not accounted for in
the derivation of Ref.~\cite{us}.

\begin{figure}
\begin{center}
\includegraphics[scale=0.47]{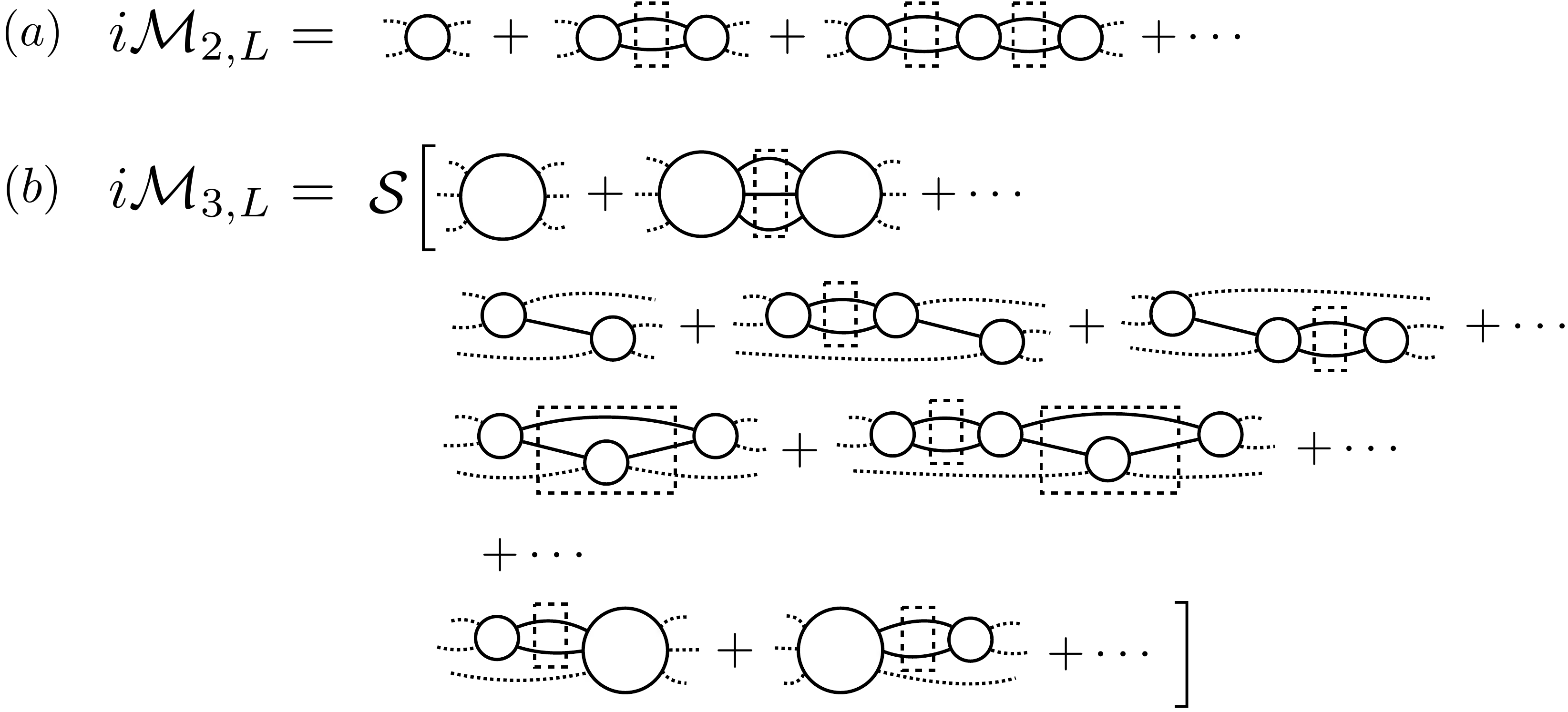}
\caption{
(a) The finite-volume two-to-two
    scattering amplitude, $\ML$.
    We restrict attention to
    two-particle CM energies below four-particle threshold,
    $E_{2}^*<4m$. As a result, only two-particle states can go
    on-shell and thus only loops shown explicitly in the figure have
    power-law finite-volume effects. The insertions between loops are
    two-to-two Bethe-Salpeter kernels, $B_2$, evaluated in
    infinite-volume. As explained in Ref.~\cite{us}, summing these
    diagrams leads to the result given in Eq.~(\ref{eq:M2toK2}).
(b) The finite-volume three-to-three amplitude,
    $\MthL$, defined as for $\ML$ but with six rather than four external
    legs. This contains both $B_2$ and the three-to-three kernels, $B_3$,
    which can be evaluated in infinite volume.
    ${\cal S} $ indicates symmetrization. 
    In both (a) and (b), dashed lines attached to the kernels indicate amputated, on-shell propagators.
    Remaining notation as in Fig.~\protect\ref{fig:fullskelexpansion}.}
\label{fig:FVamplitudes}
\end{center}
\end{figure}

Returning to $\ML$, it is useful to have a diagrammatic definition in addition to the expression in terms of $\K$ and $F$,
Eq.~(\ref{eq:M2toK2}).
This is given in Fig.~\ref{fig:FVamplitudes}(a), and defines $\ML(E_{2,k}^*, \hat a'^*, \hat a^*)$. 
Note that the spectator particle is not shown---it contributes  an additional factor of $\delta_{kk'}$.
What appears in Fig.~\ref{fig:FVamplitudes}(a)
is the same sum of two-to-two connected Feynman diagrams as occurs in the infinite-volume
two-particle scattering amplitude (with external propagators amputated and on shell)
except that spatial components of loop momenta are summed rather than integrated. 
These sums can be replaced by integrals within the Bethe-Salpeter kernels (making only exponentially suppressed corrections),
but not within the ``boxed" loops.

We now encounter for the first time an issue that will recur repeatedly in the following.
This is that the external momenta in quantities such as $\ML$ must be allowed 
to differ from 
finite-volume momenta (which we refer to as ``not being in the finite-volume set").
By contrast, internal momenta are in the finite-volume set.
This distinction is already present to some extent in Ref.~\cite{us}, 
since when the non-spectator pair are put on shell 
their directions $\hat a^*$ and $\hat a'^*$ are maintained, but
their magnitudes are changed to $q_k^*$, so that they are no longer finite-volume momenta.
Here we extend this by allowing $\hat a^*$ and $\hat a'^*$ to be arbitrary.
This allows us to decompose the angular dependence in spherical
harmonics and define the matrix form of $\ML$ used in Eq.~(\ref{eq:FthdefML}):
\begin{equation}
\label{eq:M2Langular}
4 \pi Y_{\ell' m'}^*(\hat a'^*) \mathcal M_{2,L; k'\ell'm'; k\ell m} 
      Y_{\ell m}(\hat a^*) 
\equiv \delta_{k'k} \ML(E_{2,k}^*, \hat a'^*, \hat a^*) \,.
\end{equation}
Here and in the following there is an implicit sum over repeated indices.
Note that $\ML$ is not diagonal in its angular-momentum indices,
unlike its infinite-volume counterpart, because rotation symmetry
is broken by the restriction of internal momenta to the finite-volume set.

In fact, we will need to consider more general external momenta
when we take the infinite-volume limit in Sec.~\ref{sec:limit}.
This is because we take this limit holding
the total momentum $\vec P$, as well as the external momenta of 
individual particles such as $\vec k$, fixed. 
Thus, as $L$ varies, $\vec P$ and $\vec k$ are not, in general, finite-volume momenta.
This in turn implies that the total momentum flowing through the boxed loops in
Fig.~\ref{fig:FVamplitudes}(a), which is $\vec P-\vec k$, 
is not, in general, in the finite-volume set.
We define the extended $\ML$ by keeping the summed momentum in the set,
i.e. of the form $2\pi \vec n/L$, while the other loop momentum is not in the set.
Since the Bethe-Salpeter kernels are symmetric under particle interchange, the
choice of which momentum is summed is irrelevant.

As noted above, the name ``finite-volume scattering amplitude'' is an imperfect
moniker for $\ML$. For one thing, there are no in- or out-states
in finite volume, and for another, $\ML$ is purely real (with no unitary cut).
Nevertheless, we persist with the name because, as
described in Sec.~\ref{sec:limit}, if we take the infinite-volume limit
in an appropriate way, $\ML$ does go over to the infinite-volume
scattering amplitude $\M$.

\bigskip

In the following section we derive a result closely related
to Eq.~(\ref{eq:corrresult}) in which the left-hand side is replaced with
a quantity that we call the finite-volume three-to-three scattering
amplitude, and denote $\MthL$. In direct analogy to $\ML$, $\MthL$
is defined as the sum of all on-shell, amputated six-point diagrams
evaluated in finite-volume, as shown in Fig.~\ref{fig:FVamplitudes}(b). 
As for $\ML$, the external momenta are not, in general, constrained by the finite
volume, while some of the internal momenta are. The details are described
in the following section.
The quantities on which $\MthL$ depends are simply the degrees of
freedom for two copies of three-particle phase space at fixed total
energy and momentum, i.e. the dependence is 
$\MthL(\vec k', \hat a'^*, \vec k, \hat a^*)$.  
This is closely related to the dependence
of $\ML(E_{2,k}^*, \hat a'^*, \hat a^*)$ except that now all three
particles are involved in scattering, so the partitioning into
``spectators'' ($\vec k$ and $\vec k'$) and ``scattering pairs''
($\hat a'^*$ and $\hat a^*$) is purely a choice of kinematic variables.
In particular, it is no longer true that $\vec k'=\vec k$, so that
the CM energies and momenta of the scattering pair are different
in the initial and final states.
Nevertheless, one can separately boost to the scattering-pair CM frames,
and define the directions of one of the pair as $\hat a^*$ and $\hat a'^*$
for initial and final states, respectively.
This completely fixes the kinematic configuration.

A key observation is that 
the diagrams contributing to $\MthL$, shown in Fig.~\ref{fig:FVamplitudes}(b),
are closely related to those building up $C_L$, shown in
Fig.~\ref{fig:fullskelexpansion}.
Indeed, if we first discard the 
first diagram on the first line and the entire 
second line of diagrams in the latter figure,
and then, for each remaining diagram, 
remove the $\sigma$ and $\sigma^\dagger$ blobs, amputate
and put on shell the six external propagators,
we obtain {\em exactly} the diagrams for $\MthL$.
This observation allows us to make substantial use of the work
of Ref.~\cite{us}.

As with $\ML$, calling $\MthL$ a scattering amplitude for finite $L$
is somewhat misleading, as it has no direct relation to scattering.
However, as we discuss in Sec.~\ref{sec:limit},
$\MthL$ does go over to the infinite-volume scattering amplitude $\Mth$
when $L\to\infty$, as long as one takes the limit
in the appropriate way. Thus, if we can derive a relation between
$\MthL$ and $\Kdf$, then taking the $L\to\infty$ limit gives the
desired relation between $\Mth$ and $\Kdf$.

\section{$\MthL$ with $B_2$ kernels only}
\label{sec:B2}

In this section and the next we determine the relation between
$\MthL$ and $\Kdf$. This section includes only
the $B_2$ kernels in the skeleton expansions, while the next completes the
job by adding $B_3$ kernels. 
This splitting mirrors the breakup of the derivation given in Ref.~\cite{us}.

The contributions containing only $B_2$ kernels are those of all
but the first and last lines of Fig.~\ref{fig:FVamplitudes}(b).
We find it convenient to work first with an unsymmetrized sum of these
diagrams, which we label\footnote{%
For brevity we have chosen the outgoing momenta to be $\vec p$
rather than $\vec k'$. The definitions of $\hat a'^*$ and $\hat a^*$
are as for $\MthL$ in the previous section.}
$\MthL^{(u,u)[B_2]}(\vec p, \hat a'^*; \vec k,\hat a^*)$. 
This quantity is defined by requiring that $\vec p$ and $\vec k$ are
the momenta that are not scattered by the outermost $B_2$ insertions. 
The superscript ``$u$'' indicates ``unscattered'', and ``$(u,u)$''
means that this choice of coordinates is made in both initial and final
states. This is the same notation as used in Ref.~\cite{us}.
We decompose this function in harmonics in our standard way
\begin{equation}
\label{eq:diffMthLcoords}
\MthL^{(u,u)[B_2]}(\vec p, \hat a'^*; \vec k,\hat a^*) =
4 \pi Y^*_{\ell' m'}(\hat a'^*) \mathcal
M^{(u,u)[B_2]}_{3,L;p\ell'm';k\ell m} Y_{\ell m}(\hat a^*) 
\,.
\end{equation}
Note that $\MthL^{(u,u)[B_2]}$ is defined for any $\vec p$ and $\vec k$,
but in this equation we are restricting these to be finite-volume momenta. 
 
In Ref.~\cite{us}, three-particle quantities similar to $\MthL^{(u,u)[B_2]}$
are needed only when $\vec P$, $\vec p$ and $\vec k$ are finite-volume momenta.
This implies that all three momenta in internal three-particle cuts in 
Fig.~\ref{fig:FVamplitudes}(b) can simultaneously be in the finite-volume set.
However, as noted in the previous section, when we take the infinite-volume limit
we must consider general values of the external momenta. It is then
not possible that all internal momenta are in the finite-volume set, and we
must generalize the definition of $\MthL^{(u,u)[B_2]}$. 
This is straightforward for the diagrams in the top line of 
Fig.~\ref{fig:FVamplitudes}(b): we simply take the two momenta that are being
summed to be in the finite-volume set, implying that the third is not.
Since $B_3$ is symmetric, it does not matter which two momenta are chosen.
In diagrams involving a boxed two-particle loop, we proceed as for $\ML$,
and choose (arbitrarily) one of the momenta to be in the finite-volume set.
For the remaining figures, such as those in the third line of 
Fig.~\ref{fig:FVamplitudes}(b), one has a non-trivial choice as to which of the
loop momenta to sum. However, in all cases,
the derivation of Ref.~\cite{us} provides a natural choice.
The convention we use to define the extended $\MthL^{(u,u)[B_2]}$ is
that the summed momenta always lie in the finite-volume set. 
We emphasize that the details of this choice are not relevant for our final result, 
since the distinction completely vanishes in the infinite-volume limit.

\bigskip

To identify the desired expression for $\MthL$ we begin by introducing
the relevant building blocks as presented in Ref.~\cite{us}. 
The first of these is\footnote{%
In Ref.~\cite{us} we denoted this quantity $\big[{\cal A}\big]$,
whereas here we drop the brackets.
Note that $1/(2\omega L^3)$ commutes with $\K$ and $F$,
while $\K$ and $F$ do not commute.}
\begin{align}
\mathcal A &\equiv 
\frac{i F}{2 \omega L^3} \frac{1}{1 - i \K i F} 
= \frac{1}{1 - i F i\K} \frac{i F}{2 \omega L^3} 
\,.
\label{eq:Adef}
\end{align}
This quantity is related to $\ML$, given in Eq.~(\ref{eq:M2toK2}),
by
\begin{equation}
\label{eq:MLdecom}
i \ML =  i \K + (2\omega L^3) i \K \mathcal A i \K \,.
\end{equation}
Thus $\mathcal A$ contains the volume dependence of $\ML$,
entering through the factors of $F$.

\begin{figure}
\begin{center}
\includegraphics[scale=0.18]{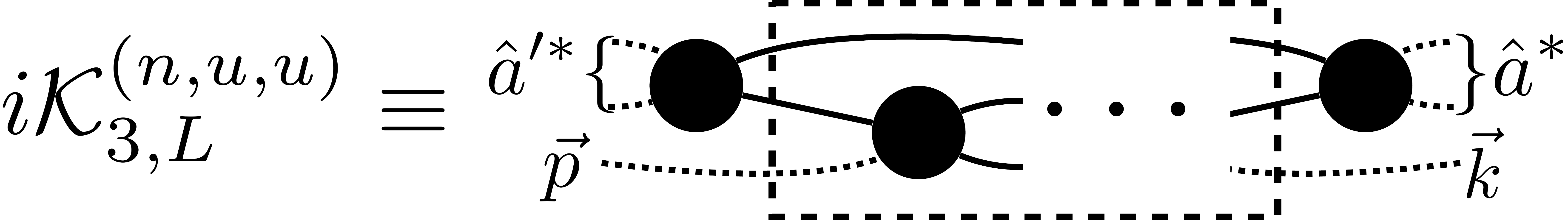}
\caption{Diagrammatic definition of  
\(\mathcal K_{3,L}^{(n,u,u)}\).
Filled circles indicate K-matrices $\K$, of which there
are $n$. Solid lines are dressed propagators, the spatial
momenta of which are summed (while time components are integrated).
Dashed lines indicate amputated, on-shell propagators.
Kinematic variables are defined as for $\MthL^{(u,u)[B_2]}$.
}
\label{fig:KthnL}
\end{center}
\end{figure}

The second building block from Ref.~\cite{us} is $\mathcal K_{3,L}^{(n,u,u)}$,
a quantity defined in Fig.~\ref{fig:KthnL}.
This object is built from $n\ge 2$ factors of $\K$, 
with adjacent insertions scattering different pairs.
It arises from sections of diagrams where
$B_2$ kernels change at least once to a different pair of propagators.
It contains power-law finite-volume effects, due to the presence of sums
over the spatial momenta in loops. Summing over all possible numbers of $\K$ factors 
leads to the quantity
\begin{equation}
 \mathcal K_{3,L}^{(u,u)} \equiv \sum_{n=2}^\infty \mathcal
 K_{3,L}^{(n,u,u)} \,.
\end{equation}
Both $\mathcal K_{3,L}^{(n,u,u)}$ and $\mathcal K_{3,L}^{(u,u)}$
can be expressed as matrices in the same fashion as $\MthL^{(u,u)[B_2]}$.

We can now state our result for the partial decomposition of
$\MthL^{(u,u)[B_2]}$:
\begin{equation}
\label{eq:MthLA}
i \MthL^{(u,u)[B_2]} = \frac{1}{1 - i \K i F} \sum_{n=0}^\infty
\left(i \mathcal K_{3,L}^{(u,u)} \mathcal A \right)^n i
\mathcal K_{3,L}^{(u,u)} \frac{1}{1 - i F i \K} \,.
\end{equation}
This decomposition makes explicit all finite-volume dependence
arising from sum-integral differences (i.e. factors of $F$)
on chains in which scattering occurs on the same pair of propagators.
It is a partial decomposition, however, because 
$\mathcal K_{3,L}^{(u,u)}$ still depends on $L$.
One can understand the result (\ref{eq:MthLA}) qualitatively as follows:
In order to have a connected scattering diagram, there must be
at least one $\mathcal K_{3,L}^{(u,u)}$ present.
This can be ``decorated'' with chains of two-to-two scattering K-matrices
and $F$ factors on either the outside or between factors of 
$\mathcal K_{3,L}^{(u,u)}$. Summing the geometric series of two-to-two chains and 
then summing over any number of $\mathcal K_{3,L}^{(u,u)}$ insertions leads to the result shown.

If the external momenta are in the finite-volume set, then
a derivation of Eq.~(\ref{eq:MthLA}) can be given by a straightforward application
of the methods of Ref.~\cite{us}. 
We provide this derivation first, and then comment (toward the end of this section)
on the generalization to arbitrary external momenta.

The most direct path to a derivation of Eq.~(\ref{eq:MthLA})
is to start from Eq.~(193) of Ref.~\cite{us}, which we reproduce here:
\begin{equation}
\label{eq:CLB2res}
C_L^{[B_2]} = C^{[B_2]}_\infty + \delta C^{[B_2]}_\infty + A'^{[B_2]}
\left[ -\frac23\frac{iF}{2\omega L^3} + \mathcal A 
  \sum_{n=0}^\infty 
\left(i \mathcal K_{3,L}^{(u,u)} \mathcal A \right)^n \right] A^{[B_2]} \,.
\end{equation}
This is an expression for the finite-volume correlator 
including only $B_2$ kernels.
Now, the main difference between $C_L$ and $\MthL$ is
the presence of the ``endcaps'' $\sigma$ and $\sigma^\dagger$ in the former.
When dressed by $B_2$ kernels these give the factors of
$A^{[B_2]}$ and $A'^{[B_2]}$ in Eq.~(\ref{eq:CLB2res}),
which are infinite-volume quantities.
To obtain $\MthL$ we can discard
these endcap factors by noting that the kinematic factor $F$,
which contains the finite-volume dependence arising
from a three-particle cut,
amputates and puts on shell the adjacent matrices.
Thus the factor between the outermost $F$ cuts in $C_L$ is
precisely the amputated on-shell $\MthL$ that we are after, as long
as the contribution contains two such cuts. 
There are no such cuts in the first two terms on the right-hand side
of Eq.~(\ref{eq:CLB2res}), since these are infinite-volume quantities,
and so they do not contribute to $\MthL$.
The third term also does not contribute since it contains only a single 
$F$ cut. 
We conclude that we need only keep the final term, i.e.
\begin{equation}
\label{eq:CLB2tokeep}
A'^{[B_2]} \left[\mathcal A 
\sum_{n=0}^\infty \left(i \mathcal K_{3,L}^{(u,u)} \mathcal A \right)^n \right]
A^{[B_2]}\,.
\end{equation}
If we drop the external factors of $A'^{[B_2]}$ and $A^{[B_2]}$ and
then further multiply by an inverse of $iF/(2 \omega L^3)$ on each
side, we are left with all amputated six-point diagrams. This includes,
however, disconnected diagrams (arising from the diagrams on the
second line of Fig.~\ref{fig:fullskelexpansion}).
Since these are not contained in the
definition of $\MthL$ they must be discarded. 
This is effected by summing from $n=1$ rather than $n=0$. 
In this way, and using the results in Eq.~(\ref{eq:Adef}) 
for $\mathcal A$, we obtain Eq.~(\ref{eq:MthLA}).

\bigskip
In the remainder of this section we follow the method of analysis used
in Sec.~IVD of Ref.~\cite{us},
where the result (\ref{eq:CLB2res}) for $C_L^{[B_2]}$ is completely
decomposed into finite- and infinite-volume quantities.
Here we apply the method to the result (\ref{eq:MthLA}) for
$\MthL^{(u,u)[B_2]}$.
In this case it turns out that we cannot simply 
borrow results from Ref.~\cite{us},
but need to introduce two new quantities.
Thus we provide a more detailed derivation.

The starting point is to use the result for ${\cal K}_{3,L}^{(u,u)}$
presented in Eq.~(215) of Ref.~\cite{us}, namely
\begin{equation}
\label{eq:KLsub}
i{\cal K}_{3,L}^{(u,u)} = iT iG [2\omega L^3] i\K +
\left(\begin{array}{cc} 1 & iT iF \end{array}\right) 
\Bigg(i\Kdf\Bigg) \sum_{j=0}^\infty \left\{ \left(\begin{array}{c} 0
  \\ 1 \end{array}\right) \frac{iF}{2\omega L^3}
\left(\begin{array}{cc} 1 & iT iF \end{array}\right) 
\Bigg(i\Kdf \Bigg) \right\}^j
\left(\begin{array}{c} 1 \\ \frac{iF}{2\omega L^3}
  iT 2\omega L^3
\end{array}\right)
\,,
\end{equation}
where [Eqs.~(212) and (213) of Ref.~\cite{us}]
\begin{align}
i T & \equiv \frac{1}{1 - i \K i G} i \K \,. \\[5pt]
\label{eq:Kdfmat}
\Bigg(i\Kdf\Bigg) & \equiv \left(\begin{array}{cc}
  i\Kdf^{(u,u)}\ \ \ \ \ & i\Kdf^{(u,s)}+i\Kdf^{(u,\tilde s)}
\\ i\Kdf^{(s,u)}+i\Kdf^{(\tilde s,u)}\ \ \ \ \ \ \ \ &
   i\Kdf^{(s,s)}+i\Kdf^{(s,\tilde s)} + i\Kdf^{(\tilde s,s)}
  +i\Kdf^{(\tilde s,\tilde s)}
\end{array}\right)\,, 
\end{align}
The result (\ref{eq:KLsub}) separates finite-volume dependence---contained in
$F$ and $G$---from the infinite-volume quantities $\K$ and $\Kdf$. 
All these quantities are matrices with our usual $k\ell m$ indices.
In Eq.~(\ref{eq:Kdfmat}),
$\Kdf$ appears in various forms, with different superscripts,
and it is convenient to collect these into a second level of
matrices as shown.
The differences between the forms of $\Kdf$ are explained
thoroughly in Ref.~\cite{us}, and here we give only an overview.

The simplest quantity to understand is $\Kdf^{(u,u)}$.
This is given by $\mathcal K_{3,L}^{(u,u)}$ except for two changes:
sums are replaced by integrals,
and singular terms are subtracted 
(along the lines discussed in the introduction)
so that it is divergence-free. 
The integrals use our non-standard pole prescription,
and, in the case of multiple loops, 
must be ordered in the manner described in Ref.~\cite{us}. Since loop momenta are integrated, there are no subtleties defining
$\Kdf^{(u,u)}$ for arbitrary external momenta. We stress that $\Kdf^{(u,u)}$ is not a physical quantity, 
because the symmetry between
the three external particles is violated by our choice of coordinates.

To define the other quantities appearing in Eq.~(\ref{eq:Kdfmat}) it
is necessary to change from the matrix form of $\Kdf^{(u,u)}$, which
decomposes angular dependence into spherical harmonics, into a form
with explicit dependence on the external momenta.
This is achieved by
\begin{equation}
\label{eq:diffKdfcoords}
\mathcal K_{\mathrm{df},3}^{(u,u)} (\vec p, \hat a'^* ; \vec k, \hat a^*) 
=
4 \pi Y^*_{\ell' m'}(\hat a'^*) \mathcal
K_{\mathrm{df},3;p\ell' m';k\ell m}^{(u,u)} Y_{\ell m}(\hat a^*) \,.
\end{equation}
The other versions of $\Kdf$
are then obtained by applying coordinate changes to this new form of $\Kdf^{(u,u)}$. 
We describe this explicitly for the example of $\Kdf^{(u,s)}$.
Recall that, in the initial state,
$\vec k$ is the momentum of the particle chosen as spectator
(so that the first two-particle kernel $B_2$ scatters the other two particles)
while $\hat a^*$ is the direction of one of the other two 
particles in their CM frame. 
In the original frame the momentum of this
second particle is called $\vec a$---a momentum that is
fully determined given $\vec k$ and $\hat a^*$ (along with $E$ and $\vec P$).
The third particle thus has momentum $\vec P-\vec k-\vec a$ in the
original frame.
A similar notation with $\vec k \to \vec p$ 
and $\vec a \to \vec a'$ is used for the final state.
As noted above, the first superscript of $\Kdf$
is related to the choice of momentum
variables for the final state, and the second to the choice for the
initial state. Focusing on the initial state,
quantities with superscripts ``$s$'' and ``$\tilde s$'' are obtained
by choosing the spectator momentum to be $\vec a$ and $\vec P-\vec k-\vec a$,
respectively. For example,
\begin{equation}
\label{eq:Kdfusdef}
\mathcal K_{\mathrm{df},3}^{(u,s)}(\vec p, \hat a'^*; \vec k, \hat a^*)
\equiv 
\mathcal K_{\mathrm{df},3}^{(u,u)}(\vec p, \hat a'^*; \vec a, \hat k^*) 
\,,
\end{equation}
where $\hat k^*$ is the direction of the particle with momentum $\vec k$
after it is boosted to the CM frame containing it and the particle
with momentum $\vec P-\vec k-\vec a$.
Note that we could equally well have used the direction of the 
particle with momentum
$\vec P-\vec k-\vec a$ after boosting in place of $\hat k^*$
as the final argument on the right-hand side of Eq.~(\ref{eq:Kdfusdef}), 
since $\Kdf^{(u,u)}$ is symmetric under this interchange.
Independently making analogous changes of variables in the final state we 
obtain all the versions of $\Kdf$.
We stress that this argument relies on $\Kdf^{(u,u)}$ being well defined for
external momenta which are not in the finite-volume set, since, even if
$\vec k$ and $\vec P$ are in this set, $\vec a$ is not.

The final step to obtain the matrix forms of $\Kdf$ appearing in Eq.~(\ref{eq:Kdfmat})
is to decompose the angular dependence in spherical harmonics.
For example, the matrix form of $\Kdf^{(u,s)}$ is obtained from the left-hand side
of (\ref{eq:Kdfusdef}) in exactly the same way as for $\Kdf^{(u,u)}$
[see Eq.~(\ref{eq:diffKdfcoords})].

As is the case for $\Kdf^{(u,u)}$, all the other forms of $\Kdf$ are
separately unphysical. To obtain a physical quantity we must symmetrize
over the initial and final assignments of momenta. 
In our matrix notation, this is achieved by
\begin{equation}
\left(\begin{array}{cc} 1&1 \end{array}\right) 
\Bigg(i\Kdf \Bigg)  
\left(\begin{array}{c} 1\\1 \end{array}\right)
 \,.
\end{equation}
Note that Eq.~(\ref{eq:KLsub}) does {\em not} contain the
symmetrized version of $\Kdf$.
A major part of the analysis in Ref.~\cite{us} 
involves rewriting $C_L^{[B_2]}$ in terms of the symmetric 
quantity, and the same is true in the following.

\bigskip

To proceed, we substitute Eq.~(\ref{eq:KLsub}) into Eq.~(\ref{eq:MthLA}) and
organize the result in powers of $\Kdf$. For example, the
contribution which is independent of $\Kdf$ is given by substituting
\begin{equation}
\label{eq:KLsub2}
i{\cal K}_{3,L}^{(u,u)} =i T iG [2\omega L^3] i\K + \mathcal O(\Kdf)
\,,
\end{equation}
[i.e. the first term in Eq.~(\ref{eq:KLsub})] to find
\begin{equation}
i \MthL^{(u,u)[B_2]} \equiv i \DL^{(u,u)} + \mathcal O(\Kdf) \,,
\end{equation}
where
\begin{equation}
i \DL^{(u,u)} =
\frac{1}{1 - i \K i F} \sum_{n=0}^\infty \left(i
T iG [2\omega L^3] i\K \mathcal A \right)^n i T 
iG [2\omega  L^3] i\K \frac{1}{1 - i F i \K}
\,.
\end{equation} 
After some algebra using the definitions of ${\cal A}$
and ${T}$ this simplifies to
\begin{equation}
\label{eq:DLuudef}
i \DL^{(u,u)} \equiv \frac{1}{1 - i \ML i G} i\ML iG i\ML [2 \omega L^3] \,.
\end{equation}
The superscript on $\DL^{(u,u)}$ is a reminder that the external momentum
indices are those of the particles not scattered by the outermost
factors of $\ML$. 

Next we consider the term linear in $\Kdf$.
To identify this it is necessary to substitute
the $\Kdf$-independent and $\mathcal O(\Kdf)$ terms 
from $i{\cal K}_{3,L}^{(u,u)}$. 
Summing all contributions to the left and right
of the single $\Kdf$ insertion then gives 
[see also Eq.~(224) of Ref.~\cite{us}]
\begin{equation}
i \MthL^{(u,u)[B_2]} \supset  \LL \Bigg(i\Kdf\Bigg) \RL  \,,
\end{equation}
where
\begin{align}
\LL & \equiv   
\left (\begin{array}{cc} 1 & 0 \end{array}\right) +  
\frac{1}{1 - i \ML i G} i \ML i F 
\left(\begin{array}{cc} 1 & 1 \end{array}\right) \,, 
\label{eq:LuLdef}
\\
\RL & \equiv  \left(\begin{array}{c} 1 \\ 0 \end{array}\right)   
+ \left(\begin{array}{c} 1 \\ 1 \end{array}\right)
\frac{i F}{2 \omega L^3} \frac{1}{1 - i \ML i G}  i \ML(2 \omega L^3)\,.
\label{eq:RuLdef}
\end{align}
These are the two new quantities that do not appear in Ref.~\cite{us}.
The superscripts again indicate a particular choice of 
spectator momentum.

To see the all orders pattern, it is sufficient to work out the 
$\mathcal O(\Kdf^2)$ term. We find this to be
\begin{equation}
\label{eq:Kdfpowertwo}
i \MthL^{(u,u)[B_2]}    
\supset \LL \Bigg(i\Kdf\Bigg) \left[\FthM + \rhoM \right]
\Bigg(i\Kdf\Bigg) \RL \,,
\end{equation}
where we have introduced two new ``second-level'' matrices,
\begin{align}
\FthM  & \equiv \left(\begin{array}{c} 1 \\ 1 \end{array}\right) 
i F_3 
\left (\begin{array}{cc} 1 & 1 \end{array}\right) \,, 
\label{eq:F3matrix}
\\
\rhoM & \equiv \left(\begin{array}{c} 1 \\ 1 \end{array}\right)  
\frac{i\rho}{2\omega }
\left(\begin{array}{cc} \frac23 & -\frac13 \end{array}\right) \,.
\end{align}
The factor between the $\Kdf$s in Eq.~(\ref{eq:Kdfpowertwo}) is
exactly the same as that in the corresponding analysis in Ref.~\cite{us}
[see Eq.~(230) of that work].
As in Ref.~\cite{us}, the term containing $\rho$ 
is integrated, rather than summed, over spectator momentum
(with sums over angular momentum indices remaining).
More precisely, our  notation means
\begin{equation}
\Bigg(i\Kdf\Bigg)
\rhoM 
\Bigg(i\Kdf\Bigg) \equiv \int \! \! \frac{d^3 s}{(2 \pi)^3} \Bigg(i\Kdf(\vec p, \vec s\,)\Bigg)
\left(\begin{array}{c} 1 \\ 1 \end{array}\right) 
\frac{i\rho(\vec s \, )}{2\omega_s}
\left(\begin{array}{cc} \frac23 & -\frac13 \end{array}\right)  
\Bigg(i\Kdf(\vec s, \vec k)\Bigg)
 \,,
 \label{eq:notabuse}
\end{equation}
where $\rho(\vec k)$ is the phase-space factor 
defined in Eq.~(\ref{eq:rhodef}), and $\vec p$ and $\vec k$ on the
right-hand side are finite-volume momenta which match the ``outside''
indices on the left hand side. The second level matrices involving $\Kdf$ on the
right hand side are the same as those defined in Eq.~(\ref{eq:Kdfmat}),
except here we replace the spectator momentum indices 
with continuous momenta. We comment that the asymmetric form in which $\rho(\vec k)$ appears here arises from our convention for ordering the $\PV$ integrals, as is described in detail in Ref.~\cite{us}.

Extending to all orders in $\Kdf$, and rearranging the sum, we deduce
\begin{equation}
i \MthL^{(u,u)[B_2]} 
= i \DL^{(u,u)} +  \LL \Bigg (i \Kdf^{\mathrm{[B_2, \rho]}} \Bigg) 
\sum_{n=0}^\infty 
\left [\FthM \Bigg ( i \Kdf^{\mathrm{[B_2 ,\rho ]}} \Bigg) \right ]^n
   \RL \,,
\label{eq:MthLuures}
\end{equation}
which is the analog of Eq.~(234) of Ref.~\cite{us}.
Here the $\rho$-dependent terms are summed into the new infinite-volume
quantity
\begin{equation}
\label{eq:KdfBtworho}
\Bigg(i\Kdf^{[B_2,\rho]}\Bigg)  \equiv   \sum_{n=0}^\infty    \Bigg(i\Kdf\Bigg)
\left [\rhoM
\Bigg(i\Kdf\Bigg) \right ]^n \,.
\end{equation}
We refer to these $\rho$-dependent terms as ``decorations''.

The result (\ref{eq:MthLuures}) has been derived assuming that the external
spectator momenta and the total momentum take finite-volume values.\footnote{%
The momenta of the scattered pair in the initial and final states are 
not in the finite-volume set, due to the on-shell condition, as discussed 
around Eq.~(\ref{eq:M2Langular}).
The direction vectors for these momenta, $\hat a^*$ and $\hat a'^*$, can be
chosen arbitrarily.
}
We now argue that it holds for arbitrary choices of these momenta,
as long as the finite-volume quantities $F$ and $G$ contained in $F_3$ and
$\ML$ are extended to arbitrary external momenta in the very straightforward
manner described in the appendix.
The derivation of Eq.~(\ref{eq:MthLuures}) relies
on repeated replacements of momentum sums over three particle cuts
with momentum integrals plus the  difference, $F$.\footnote{%
To see this most clearly, one should expand out the geometric series contained
in $F_3$, $\ML$ and ${\cal D}_L^{(u,u)}$, in which case ${\cal M}_{3,L}^{(u,u)[B_2]}$
is a sum of terms each of which
consists of products of either $F$ or $G$ alternating
with factors of $\K$ and $\Kdf^{[B_2,\rho]}$.} 
One must ensure that replacements are done in an appropriate
order (and that integrals are defined with the $\PV$ prescription), as discussed
in detail in Ref.~\cite{us}.
But what matters here are the following observations. 
First,
the determination of which summands lead to a non-trivial sum-integral 
difference (i.e. the three particle cuts) is independent of the choice of
external momenta. 
Second, the difference function $F$ maintains its form (given in the appendix)
for arbitrary external momenta.
Third, the sums over spectator momentum indices,
implicitly contained in the multiplication of matrices in Eq.~(\ref{eq:MthLuures}) 
(and which are, by definition, over finite-volume momenta)
correspond precisely to those loop momenta which should be summed
over finite-volume momenta,
according to the diagrammatic definition of $\MthL^{(u,u)[B_2]}$ given at the beginning of
this section. In other words, the correct choice of which internal momenta to keep in
the finite-volume set is being made.
And, finally, the kinematic factor $G$ (which contains no momentum sums)
extends to non-finite-volume momenta in a very straightforward way, as described
in the appendix.
Combining these observations, it is straightforward to see that the expression
for $\MthL^{(u,u)[B_2]}$ given above remains valid for arbitrary external momenta,
with one proviso.
The proviso is that the external momenta, which propagate into the expression until there is a
factor of $G$ (either in ${\cal D}_L^{(u,u)}$, ${\cal L}_L^{(u)}$, or ${\cal R}_L^{(u)}$)
or a factor of $\Kdf^{[B_2,\rho]}$, are not in the finite-volume set. 
Thus the first matrix that appears
on either end of the expression has different indices from matrices in the middle of the expression.
In particular, the matrices on the end have one index which is an
external momentum while the other is an (internal) finite-volume momentum.
Fortunately, these subtleties become irrelevant when one takes the infinite-volume limit
and sums become integrals.

\bigskip

The final step in the construction of the contribution of 
$\MthL$ from $B_2$ kernels is to symmetrize over
initial and final momentum assignments.
However, a complication arises because 
${\cal D}_L^{(u,u)}$, ${\cal L}_L^{(u)}$ and ${\cal R}_L^{(u)}$
contain poles.
When one changes the assignment of momentum labels, 
the angular dependence now sweeps over the poles.
Since this implies that the functions are  not square integrable,
one cannot decompose into spherical harmonics.
To avoid this technical problem one must first 
change from spherical harmonic indices to
an explicit dependence on angular variables, and only then symmetrize. 
For almost all choices of external momenta, one then avoids the singularity. 

To make the symmetrization operator explicit, we first convert to angular variables
for $\MthL^{(u,u)[B_2]}$ in the standard way:
\begin{equation}
\label{eq:MthLangtosph}
\MthL^{(u,u)[B_2]} (\vec p, \hat a'^* ; \vec k, \hat a^*) 
=
4 \pi Y^*_{\ell' m'}(\hat a'^*) \mathcal
{\cal M}_{3,L;p\ell' m';k\ell m}^{(u,u)[B_2]} Y_{\ell m}(\hat a^*) \,.
\end{equation}
Note that, as discussed above at length, we can treat all the external momenta
as continuous variables, not constrained to lie in the finite-volume set.
We now define the quantities with $u$ replaced by $s$ and $\tilde s$ in exactly
the same way as discussed for $\Kdf$ above---see Eq.~\ref{eq:Kdfusdef} and surrounding text.
We stress that this is simply implementing the interchange of the choice of spectator particle.
Then we have 
\begin{align}
\MthL^{[B_2]}(\vec p, \hat a'^*; \vec k, \hat a^*)
&= \sum_{x=u,s,\tilde s} \sum_{y=u,s,\tilde s}
\MthL^{(x,y)[B_2]}(\vec p, \hat a'^*; \vec k, \hat a^*)
\label{eq:symmMthL}
\\
&\equiv 
\mathcal S \Big \{  \mathcal M^{(u,u)[B_2]}_{3,L;p, \ell', m';  k, \ell, m}  \Big \} \,.
\label{eq:Sdef}
\end{align}
The second line defines the action of the symmetrization operator ${\cal S}$,
and applies to any ``$(u,u)$'' quantity that can be defined for general
external momenta.

Applying ${\cal S}$ to both sides of Eq.~(\ref{eq:MthLuures}) we find, after
some algebra, the final result of this section,
\begin{multline}
\label{eq:MthLtwres}
i \MthL^{[B_2]}(\vec p,\hat a'^*;\vec k,\hat a^*)= i \DL(\vec p,\hat a'^*;\vec k,\hat a^*)
\\
+ \mathcal S\bigg\{ 
\left[\frac13 + \frac{1}{1\! -\! i \ML i G} i \ML i F\right] 
i \Kdf^{[B_2,\rho]} \frac{1}{1\!-\! i F_3 i \Kdf^{[B_2,\rho]}} 
\left[\frac13 + \frac{i F}{2 \omega L^3}\frac{1}{1 \!-\! i \ML i G} 
i \ML (2 \omega L^3) \right ] \bigg \} \,.
\end{multline}
Here $\DL$ is the symmetrized form of $\DL^{(u,u)}$,
\begin{equation}
\DL(\vec p,\hat a'^*;\vec k,\hat a^*)
= \mathcal S \bigg\{ \mathcal D_L^{(u,u)}\bigg\}\,,
\end{equation}
and
\begin{equation}
i \Kdf^{[B_2,\rho]} \equiv 
\left(\begin{array}{cc} 1&1 \end{array}\right) 
\Bigg(i\Kdf^{[B_2,\rho]}\Bigg)  
\left(\begin{array}{c} 1\\1 \end{array}\right) \,.
\label{eq:KdfB2rhodef}
\end{equation}
equals the quantity of the same name appearing in Ref.~\cite{us}.

Several comments are in order about the derivation of
Eq.~(\ref{eq:MthLtwres}).
When symmetrizing over the row vector ${\cal L}_L^{(u)}$, defined in
Eq.~(\ref{eq:LuLdef}), the row vector $\left(1\ 0\right)$ within it
is converted into $\left(1\ 1\right)$.\footnote{%
Recall that our second-level matrix notation combines
superscripts $s$ and $\tilde s$ into the lower entry; see, e.g.,
Eq.~(\ref{eq:Kdfmat}).}
Since the final form (\ref{eq:MthLtwres})
still contains an external symmetrization operator, 
we can replace $\left(1\ 0\right)$ with $\left(1\ 1\right)$ as long
as we include a factor of $1/3$.
The row vector $\left(1\ 1\right)$ can now be factored out
and applied to the right.
A similar discussion holds for ${\cal R}_L^{(u)}$, with the
symmetric column vector $\left(\begin{array}{c} 1\\1\end{array}\right)$
factoring out at its left-hand end.
Together with the fact that the matrix form of $F_3$,
Eq.~(\ref{eq:F3matrix}), contains symmetric column and row vectors,
this implies that all entries of the matrix $\left( i\Kdf^{[B_2,\rho]}\right)$
in Eq.~(\ref{eq:MthLuures}) are now sandwiched between symmetric vectors.
This is why the symmetric quantity $\Kdf^{[B_2,\rho]}$ of
Eq.~(\ref{eq:KdfB2rhodef}) appears in our final result.
One is then left with a geometric series of powers of
$F_3 \Kdf^{[B_2,\rho]}$ which sums to the form shown in
Eq.~(\ref{eq:MthLtwres}).

We stress that $\Kdf^{[B_2,\rho]}$, while symmetric under external
particle exchange, violates this symmetry internally.
This is because of the above-mentioned decorations involving the
matrix $\rho$, which contains the asymmetric row vector
$\left(2/3, -1/3\right)$. Furthermore, our $\PV$ pole prescription
breaks particle interchange symmetry.
These facts make $\Kdf^{[B_2,\rho]}$ a difficult quantity to interpret. 

It is of course crucial that Eq.~(\ref{eq:MthLtwres}) 
contains the same $\Kdf^{[B_2,\rho]}$ as appears in the
analogous expression for $C_L^{[B_2]}$ given in Eq.~(234) of Ref.~\cite{us}.
We will see that this continues to hold when $B_3$ kernels are
included. This means that the same quantity $\Kdf$ that appears in the
quantization condition will appear in the relation to $\Mth$.

\section{$\MthL$ with $B_2$ and $B_3$ kernels}
\label{sec:B3}

In this section we complete the decomposition of $\MthL$ by including
all remaining diagrams, i.e. those with at least one $B_3$ kernel. 
This set is illustrated
by the first and last lines of Fig.~\ref{fig:FVamplitudes}(b). 
Although the final result is straightforward, 
with the form of Eq.~(\ref{eq:MthLtwres}) being preserved aside from
a new definition of $\Kdf$, the derivation of this result is quite involved.
While we are able to partly reuse results from
Ref.~\cite{us}, significant additional work is required. 
As in the previous section, we derive the result first with 
finite-volume external spectator and total momenta, and then note that
the result maintains its form if these momenta are allowed to be general.

\begin{figure}
\begin{center}
\includegraphics[scale=0.5]{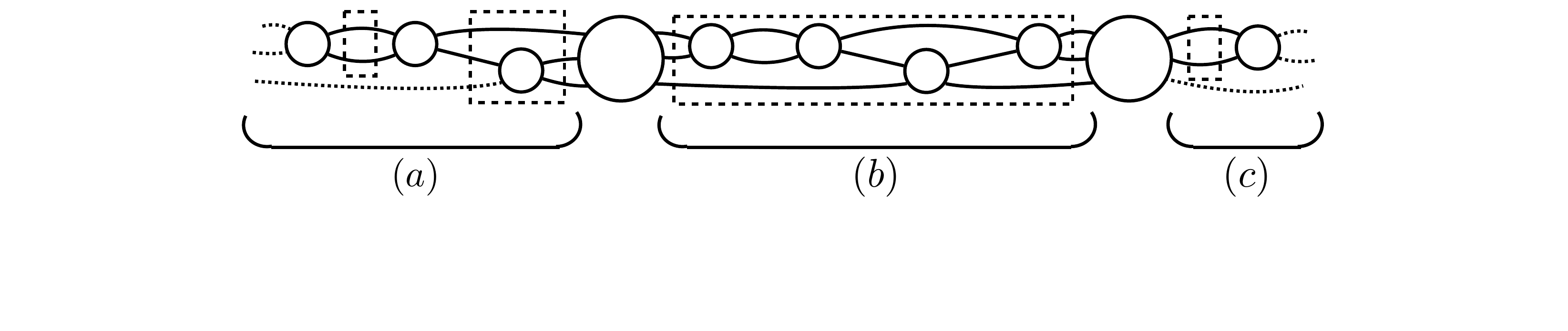}
\vspace{-40pt}
\caption{Example of the decomposition of a contribution to $\MthL$ 
in terms of segments separated by insertions of the three-to-three kernel $B_3$. 
As discussed in the text, the segments are of three types:
(a) those connecting the final state to the leftmost kernel;
(b) those which lie between two $B_3$ kernels; 
and (c) those connecting the rightmost kernel to the initial state. 
Segments of type (b) were analyzed in Ref.~\cite{us}, 
whereas those of type (a) and (c) are new to this work.}
\label{fig:figsegments}
\end{center}
\end{figure}

$B_3$ kernels are symmetric functions of incoming and outgoing momenta
that are smooth in the energy range of interest. They naturally divide
diagrams contributing to $\MthL$ into segments within which
finite-volume dependence can arise, as illustrated in Fig.~\ref{fig:figsegments}. 
These segments can be treated separately, and are of three types.
The first connects the final state
fields to the leftmost $B_3$ kernel, and appears on the left-hand end
of all diagrams considered in this section.
The second type lies between two $B_3$ kernels, and
can appear any number of times (including zero) in the middle of
the diagram.
Finally, on the right-hand end there is always a segment
connecting the rightmost $B_3$ to the initial state. 
The second (``middle'') type of segment was analyzed in Ref.~\cite{us}, 
while the other two are new to this work.

We first recall the expression for a middle segment. 
To understand this we need the final result for  finite-volume
correlator including only $B_2$ kernels, $C_L^{[B_2]}$.
This is given by Eq.~(239) of \cite{us}:
\begin{equation}
\label{eq:CLB2final}
C_L^{[B_2]} = C_\infty^{[B_2,\rho]} +
A'^{[B_2,\rho]} {\cal Z} A^{[B_2,\rho]}
\,,
\end{equation}
with
\begin{equation}
\label{eq:Zdef}
{\cal Z} \equiv \frac1{1- iF_3 i\Kdf^{[B_2,\rho]}} iF_3
\,.
\end{equation}
This is obtained by performing manipulations analogous to
those of the previous section on the starting form, Eq.~(\ref{eq:CLB2res}).
$C_\infty^{[B_2,\rho]}$ is the infinite-volume version of $C_L$ with
only $B_2$ kernels included,
using the $\PV$ prescription, and including decorations due to $\rho$.
It is defined in Sec.~IVD of Ref.~\cite{us}.
The second term in (\ref{eq:CLB2final}) contains the finite-volume
dependence, which is collected into ${\cal Z}$.
All infinite-volume contributions between the $\sigma$ at the left
and the first finite-volume cut $F$ are contained in
$A'^{[B_2,\rho]}$, while $A^{[B_2,\rho]}$ contains such contributions
between the rightmost $F$ and the $\sigma^\dagger$.
Following Ref.~\cite{us},
we define ``decoration operators'' $D_C$, $D_{A'}$ and $D_{A}$ such that
[Eq.~(241) of \cite{us}]
\begin{align}
C_\infty^{[B_2,\rho]} &\equiv \sigma  D_C^{[B_2,\rho]} \sigma^\dagger
\,,
\label{eq:DCdef}
\\
A'^{[B_2,\rho]} &\equiv \sigma\; D_{A'}^{[B_2,\rho]}
\,, 
\label{eq:DApdef}
\\
A^{[B_2,\rho]} &\equiv D_A^{[B_2,\rho]} \sigma^\dagger
\,.
\label{eq:DAdef}
\end{align}
These are infinite-volume integral operators.
The point of these definitions is that the decoration operators
do not depend on the choice of functions at the ends.
In $C_L^{[B_2]}$ these are $\sigma$ and $\sigma^\dagger$,
but if both are replaced by $B_3$ then one obtains a
middle segment of $\MthL$.
Thus the latter can be written 
\begin{equation}
\label{eq:middlesegment}
\dots i B_3 M^{[B_2,\rho]} i B_3 \dots \,, \ {\rm where}\
M^{[B_2,\rho]}= D_C^{[B_2,\rho]} + D_{A'}^{[B_2,\rho]}{\cal Z}
D_A^{[B_2,\rho]}
\,,
\end{equation}
which is Eq.~(243) of \cite{us}.

We now turn to the end segments.
We call the integral operator appearing
in the left-hand segment $L^{(u)[B_2,\rho]}$, and the corresponding operator
for the right-hand segment $R^{(u)[B_2,\rho]}$.
Using this notation, the contribution to $\MthL^{(u,u)}$ containing
one and two $B_3$s is
\begin{equation}
i\MthL^{(u,u)} \supset L^{(u)[B_2,\rho]} iB_3 R^{(u)[B_2,\rho]}
+ L^{(u)[B_2,\rho]} iB_3 M^{[B_2,\rho]} iB_3 R^{(u)[B_2,\rho]}
\,.
\end{equation}
No $(u,u)$ superscripts are necessary for the middle segment due to
the symmetry of $B_3$. 

The result for $R^{(u)[B_2,\rho]}$ can be obtained by reflection
from that for $L^{(u)[B_2,\rho]}$, so we focus on the latter.
Our starting point is a result that follows from Eq.~(174) of \cite{us}
by the same argument given above for obtaining Eq.~(\ref{eq:MthLA})
from Eq.~(\ref{eq:CLB2res}). We find
\begin{equation}
\label{eq:Lu1}
L^{(u)[B_2,\rho]} \sigma^\dagger
= \frac1{1-i \mathcal K_2 iF} 
\sum_{n=0}^\infty \left(i \mathcal K_{3,L}^{(u,u)} \mathcal A \right)^n 
\sum_{m=0}^\infty A_L^{(m,u)} - \frac23 \sigma^\dagger
\,.
\end{equation}
Note that we are acting $L^{(u)[B_2,\rho]}$ on $\sigma^\dagger$ rather
than $B_3$. This is just a convenience, as it allows us to reuse
notation from Ref.~\cite{us}. The operator $L^{(u)[B_2,\rho]}$ is
the same for either choice. The new quantity $A_L^{(m,u)}$
is defined in Ref.~\cite{us}, and represents diagrams in which
$\sigma^\dagger$ is dressed with $m$ factors of $\K$, with the
scattering pair switching between each factor. All loops except that
closest to the $\sigma^\dagger$ involve sums over spatial
momenta, and thus contain $L$ dependence. This is illustrated in
Fig.~17(a) of \cite{us}. The leading term is $A_L^{(0,u)}=\sigma^\dagger$
(which will ultimately become $B_3$ in our case).
The last term on the right-hand side in Eq.~(\ref{eq:Lu1}) is needed
since the first term equals $\sigma^\dagger$ at leading order, 
but we require $\sigma^\dagger/3$ in the unsymmetrized endcap. 
In the construction of $\MthL^{(u,u)}$ this will correspond to
 factors of $B_3$ being multiplied by $1/3$ whenever they are 
the outermost insertions. This is the correct definition to recover a factor 
of $B_3$ without the $1/3$ from such diagrams within the {\em symmetrized} amplitude. 
Since $B_3$ is itself symmetric, the effect of symmetrization on such diagrams 
is multiplication by $3$, which cancels the $1/3$ included here.  

The form (\ref{eq:Lu1}) is next rewritten using the results
in Eqs.~(188), (191) and (192) in \cite{us}, leading after some
algebra to
\begin{equation}
\label{eq:Lu2}
L^{(u)[B_2,\rho]} \sigma^\dagger
= \frac1{1-i \mathcal K_2 iF} 
\sum_{n=0}^\infty \left(i \mathcal K_{3,L}^{(u,u)} \mathcal A \right)^n 
A^{[B_2]}
+ \frac13 \sigma^\dagger + A^{(u)[B_2]}-A^{[B_2]}
\,.
\end{equation}
Here $A^{(u)[B_2]}$ and $A^{[B_2]}$ are infinite-volume quantities
involving any number of $\K$s connected to $\sigma^\dagger$.
$A^{[B_2]}$ is the symmetrized version, which has already appeared in
Eqs.~(\ref{eq:CLB2res}) and (\ref{eq:CLB2tokeep}).
Both quantities are defined in Ref.~\cite{us}.\footnote{{In Ref.~\cite{us}, $A^{(u)}$ denotes what we refer to here as $A^{(u)[B_2]}$. Note that $A^{(u)[B_2]}$ does not include an isolated factor of $\sigma^\dagger$, i.e.~all terms in $A^{(u)[B_2]}$ have at least one insertion of $\K$. By contrast the symmetrized version, $A^{[B_2]}$ does start with an isolated $\sigma^\dagger$ [$A^{[B_2]} = \sigma^\dagger + \mathcal O(\K)$].}}
The result (\ref{eq:Lu2}) is the analog of Eq.~(\ref{eq:MthLA}) above
when there is a $B_3$ on the right-hand end rather than an initial state.

Using the definitions of ${\cal A}$ and 
$\MthL^{(u,u)[B_2]}$, Eqs.~(\ref{eq:Adef}) and (\ref{eq:MthLA}) respectively,
we can rewrite our result for $L^{(u)[B_2,\rho]}$ as
\begin{equation}
\label{eq:Lu3}
L^{(u)[B_2,\rho]} \sigma^\dagger
=
i\MthL^{(u,u)[B_2]} \frac{iF}{2\omega L^3} A^{[B_2]}
+ \frac13 \sigma^\dagger + A^{(u)[B_2]} 
+ \frac{i\K i F}{1-i\K iF} A^{[B_2]}
\,.
\end{equation}
This allows us to reuse the work from the previous
section, which led to the result (\ref{eq:MthLuures})
for $\MthL^{(u,u)[B_2]}$.
Substituting the latter into Eq.~(\ref{eq:Lu3}), we encounter
a new combination that can be rewritten as follows 
\begin{align}
\RL \frac{iF}{2\omega L^3} A^{[B_2]}
&=
\left\{ 
\left(\begin{array}{c} 1\\ 1 \end{array}\right) i F_3
+
\left(\begin{array}{c} 2/3\\ -1/3 \end{array}\right) 
\frac{i\rho}{2\omega} \right\}
A^{[B_2]}
\\
&= \left\{\FthM + \rhoMT \right\} \Acol
\,,
\end{align}
where
\begin{equation}
\rhoMT \equiv  
\left(\begin{array}{c} 2/3 \\ -1/3 \end{array}\right)  
\frac{i\rho}{2\omega }
\left(\begin{array}{cc} 1 & 1 \end{array}\right) \,,
\end{equation}
and
\begin{equation}
\Acol \equiv 
\left( 
\begin{array}{c} \sigma^\dagger/3 + A^{(u)[B_2]} 
\\[5pt] 
2 \sigma^\dagger/3+A^{(s)[B_2]} + A^{(\tilde s)[B_2]}\end{array} \right) \,.
\end{equation}
Here we have split up the symmetric quantity $A^{[B_2]}$ into
a column matrix, such that
\begin{equation}
A^{[B_2]} = \left(\begin{array}{cc} 1 & 1 \end{array}\right) \Acol
\,.
\end{equation}
Introducing the definition [equivalent to Eq.~(237) of \cite{us}]
\begin{equation}
   \Arhocol \equiv  \Acol + 
\Bigg(i\Kdf^{[B_2,\rho]}\Bigg)
\rhoMT \Acol \,,
\end{equation}
we find, after considerable algebra, that we can rewrite (\ref{eq:Lu3}) as
\begin{equation}
\label{eq:Lu4}
L^{(u)[B_2,\rho]} \sigma^\dagger
=
\LL \sum_{n=0}^\infty \left\{
\Bigg(i\Kdf^{[B_2,\rho]}\Bigg) \FthM \right\}^n \Arhocol
\,.
\end{equation}
Finally, defining asymmetric decoration operators by
\begin{equation}
\left( \begin{array}{c} D_A^{(u)[B_2,\rho]} 
\\[5pt] 
D_A^{[B_2,\rho]} - D_A^{(u)[B_2,\rho]}\end{array}\right )  
\sigma^\dagger \equiv \Arhocol \,,
\end{equation}
we can write a stand-alone expression for the integral operator appearing
in the left-hand segment
\begin{equation}
\label{eq:Lu5}
L^{(u)[B_2,\rho]}
=
\LL \sum_{n=0}^\infty \left\{
\Bigg(i\Kdf^{[B_2,\rho]}\Bigg) \FthM \right\}^n 
\left( \begin{array}{c} D_A^{(u)[B_2,\rho]} 
\\[5pt] 
D_A^{[B_2,\rho]} - D_A^{(u)[B_2,\rho]}\end{array}\right )  
\,.
\end{equation}
As for the middle segment, $M^{[B_2,\rho]}$,
finite-volume dependence enters through the factors of $F_3$.

The corresponding expression for the right-hand segment is
\begin{equation}
\label{eq:Ru1}
R^{(u)[B_2,\rho]}
=
\left( \begin{array}{cc} D_{A'}^{(u)[B_2,\rho]}, &
D_{A'}^{[B_2,\rho]} - D_{A'}^{(u)[B_2,\rho]}\end{array}\right )
\sum_{n=0}^\infty \left\{
\FthM \Bigg(i\Kdf^{[B_2,\rho]}\Bigg) \right\}^n 
 \RL
\,,
\end{equation}
where we have defined the mirrored decoration operators
\begin{equation}
\sigma \left( 
\begin{array}{cc} D_{A'}^{(u)[B_2,\rho]}, &  
D_{A'}^{[B_2,\rho]} - D_{A'}^{(u)[B_2,\rho]} \end{array} \right) 
\equiv 
\APrhorow \,, 
\end{equation}
with
\begin{equation}
 \APrhorow  = \AProw  + \AProw   \rhoM
    \Bigg(i\Kdf^{[B_2,\rho]}\Bigg)  \,,
\end{equation}
where
\begin{equation}
\AProw \equiv \left(
\begin{array}{cc} \sigma/3+ A'^{(u)}, \ \ &
\ \  2 \sigma/3 + A'^{(s)} +A'^{(\tilde s)} \end{array} \right ) \,.
\end{equation}
The quantities $A'^{(u)}$, $A'^{(s)}$ and $A'^{(\tilde s)}$ are defined
in \cite{us}.

We now have all the pieces needed to construct the full unsymmetrized finite-volume
scattering amplitude. The result is
\begin{equation}
i\MthL^{(u,u)} = 
i\MthL^{(u,u)[B_2]} +
L^{(u)[B_2,\rho]} iB_3 \sum_{n=0}^\infty
\left\{ M^{[B_2,\rho]} iB_3 \right\}^n\  R^{(u)[B_2,\rho]}
\,.
\end{equation}
Reorganizing the sums, and introducing the quantity
\begin{equation}
\Bigg ( i \Kdf^{\mathrm{[B_2 \& B_3]}} \Bigg)  
\equiv 
\Bigg (i \Kdf^{\mathrm{[B_2 , \rho]}} \Bigg)
+\Bigg ( i B_3^{[B_2,\rho]} \Bigg ) \,,
\end{equation}
where
\begin{equation}
\Bigg ( iB_3^{[B_2,\rho]} \Bigg ) \equiv
\sum_{n=0}^\infty 
\left( \begin{array}{c} D_A^{(u)[B_2,\rho]} \\[5pt] 
D_A^{[B_2,\rho]} - D_A^{(u)[B_2,\rho]}\end{array} \right) 
\left[i B_3 D_{C}^{[B_2,\rho]}\right]^n iB_3 
\left( \begin{array}{cc} D_{A'}^{[B_2,\rho]}, &  
D_{A'}^{[B_2,\rho]} - D_{A'}^{(\tilde u)[B_2,\rho]} \end{array} \right)
\,,
\end{equation}
we find
\begin{equation}
\label{eq:MthLuures2}
i \MthL^{(u,u)} = i \DL^{(u,u)} + 
\LL \Bigg(i \Kdf^{\mathrm{[B_2 \& B_3]}} \Bigg)
\sum_{n=0}^\infty   
\left[\FthM \Bigg(i \Kdf^{\mathrm{[B_2 \& B_3 ]}} \Bigg) \right ]^n   
\RL \,.
\end{equation}
As claimed above, this has exactly the same form as the result
(\ref{eq:MthLuures}) from the previous section, but now the $B_3$
contributions are contained within the modified $\Kdf$. As in the previous section, this result remains valid when the external momenta
take arbitrary values instead of being restricted to the finite-volume set. 
Indeed, the argument for this is unchanged, since the new features
introduced by factors of $B_3$ only impact the infinite-volume quantities in
Eq.~(\ref{eq:MthLuures2}) for which the transition to arbitrary external momenta is
not problematic.

Symmetrizing proceeds as in the previous section, and leads to
the final form for the finite-volume scattering amplitude:\footnote{We note that there is a simple mnemonic that one can use to obtain this result from that for the full finite-volume correlator given in Eq.~(\ref{eq:corrresult}). This follows the procedure used to derive Eq.~(\ref{eq:MthLA}) from the corresponding
result for the finite-volume correlator, Eq.~(\ref{eq:CLB2res}).
More precisely, identifying terms with at least two insertions of $F$, discarding the endcaps $A'$ and $A$, multiplying each side by an inverse of $iF/(2 \omega L^3)$, discarding disconnected diagrams, and then symmetrizing, one reaches the final expression for the finite-volume three-to-three scattering amplitude given in Eq.~(\ref{eq:MthLres}). This procedure cannot, however, be rigorously justified,
for two reasons. The first issue is that various redefinitions of the infinite-volume quantities were required to reach Eq. (\ref{eq:corrresult}).
This obscures which diagrams are contributing to which quantities, and the only way we could resolve this uncertainty was by starting at an earlier stage of the derivation, Eq. (\ref{eq:CLB2res}), before such redefinitions have been made. The second issue is that the quantity reached by amputating $A$, $A'$ as well as the outermost $iF/(2 \omega L^3)$ has ill-defined exchange symmetry. It contains certain contributions which are symmetrized and others which are not. It turns out that symmetrizing the result in the manner described in the text removes this ambiguity, without overcounting or neglecting terms.}
\begin{multline}
\label{eq:MthLres}
i \MthL(\vec p,\hat a'^*;\vec k,\hat a^*) = i \DL(\vec p,\hat a'^*;\vec k,\hat a^*)
\\
+ \mathcal S\bigg\{ 
\left[\frac13 + \frac{1}{1\! -\! i \ML i G} i \ML i F\right] 
i \Kdf \frac{1}{1\!-\! i F_3 i \Kdf} 
\left[\frac13 + \frac{i F}{2 \omega L^3}\frac{1}{1 \!-\! i \ML i G} 
i \ML (2 \omega L^3) \right ] \bigg \} \,.
\end{multline}
Here $\Kdf$, given by
\begin{equation}
i \Kdf \equiv 
\left(\begin{array}{cc} 1,&1 \end{array}\right) 
\Bigg(i\Kdf^{[B_2 \& B_3 ]}\Bigg)  
\left(\begin{array}{c} 1\\1 \end{array}\right) \,,
\end{equation}
is equal to the quantity of the same name appearing in 
Eq.~(251) of Ref.~\cite{us}. 
Thus we have achieved our objective---expressing $\MthL$ in terms of
the same intermediate quantity $\Kdf$ that appears in our quantization
condition.

The result (\ref{eq:MthLres}) suggests an intuitive interpretation of $\Kdf$
as an effective local three-particle interaction.
In words, one can describe the second term on the right-hand side
of Eq.~(\ref{eq:MthLres}) as having any number of two-particle
interactions, jumping back and forth between scattering pairs,
followed by a three-particle interaction (given by $\Kdf$),
followed by more two-to-two scattering ($F_3$), possibly another
three-particle interaction, etc., and ending with more two-to-two
scattering.
We note, however, that separating the local part of the three-particle
scattering process involves some ambiguity, which here is manifested
by the presence of the cut-off function $H$, whose detailed form
is arbitrary.

\section{Infinite-Volume Limit}
\label{sec:limit}

In this section we take the infinite-volume limit of
Eq.~(\ref{eq:MthLres}) and thereby derive a relation between $\Mth$,
the standard three-to-three scattering amplitude, and $\Kdf$, the
infinite-volume quantity appearing in the quantization condition.

As a warm-up, we consider first the infinite-volume limit of
the two-to-two finite-volume scattering amplitude,
whose form we recall is
\begin{equation}
\label{eq:MLres}
\ML = \frac{1}{1+ \K F} \K \,,
\end{equation}
with $F$ defined in Eqs.~(\ref{eq:Fdef1})-(\ref{eq:Fdef3}). $\ML$, $\K$ and $F$ are all matrices with both spectator momentum 
and angular momentum indices.
The total energy and momentum is $(E,\vec P)$,
while we call the incoming and outgoing spectator momenta 
$\vec k$ and $\vec p$, respectively.
As stressed repeatedly above, these momenta are not constrained to lie in the finite-volume set.
We take $L\to\infty$ holding these momenta fixed.

The crucial question when taking the infinite-volume limit is how to treat
sums over finite-volume momenta.
We know that (suitably normalized) sums go over to integrals
\begin{equation}
\frac1{L^3} \sum_{\vec p} \longrightarrow \int \! \frac{d^3p}{(2 \pi)^3}\,,
\end{equation}
up to corrections which vanish as $L\to\infty$. If the summand is
smooth the corrections vanish exponentially, while if there are
singularities the corrections fall as powers of $L$. Keeping track
of the latter was the main task in the analysis of Ref.~\cite{us}.
Here, by contrast, we do not care how
the corrections fall off, only that they vanish in the limit.
However a subtlety arises for summands containing singularities,
because the corresponding integrals are only well-defined with a pole
prescription. For any finite $L$, the finite-volume momenta will avoid
the singularities unless the energy $E$ is equal to that of three
non-interacting particles. But as $L \to \infty$, these singularities
become arbitrarily dense and the limit is ill-defined. 
To take the limit we must
first introduce a pole prescription for the sum.
The natural choice is the $i\epsilon$ prescription, 
since this is the choice used in defining scattering amplitudes
in infinite volume. We also consider below what happens if we
use the $\PV$ prescription.

This $i\epsilon$ prescription works for sums in a straightforward way.
The procedure is as follows:
(1) Replace all poles in
sums by poles with nonzero $i \epsilon$; 
(2) Send  $L \rightarrow \infty$ at fixed $i \epsilon$---this limit is 
well defined with all sums becoming integrals;
(3) Send $\epsilon \rightarrow 0$.
In this limit the contribution of each diagram in the
skeleton expansion for $\ML$ [see Fig.~\ref{fig:FVamplitudes}(a)] is converted into the corresponding
contribution for $\M$.
Thus $\ML \to \M$ when using this prescription, 
and similarly $\MthL \to \Mth$. 

Focusing first on $\ML$, we note that it contains two types of finite-volume 
sums: those arising from matrix
multiplications over the index $k$, and that contained in $F$.\footnote{%
The matrix multiplications over the angular momentum indices $\ell m$
go over unchanged to the infinite-volume limit.}
In this case (but not for $\MthL$ below) the former sums are trivial.
This is because spectator momentum is conserved,
so all matrices are proportional to $\delta_{k_1 k_2}$.
Multiplying these out leads\footnote{%
There is a minor inconsistency between our matrix notation---designed for summing over
finite-volume momenta---and the fact that we have generalized to arbitrary 
spectator momenta.
In particular, since the external momenta do not appear in the finite-volume set, what
does the $\delta_{k_1 k_2}$ mean? Working back through the derivation, we find that
it is simply a mnemonic for spectator-momentum conservation, 
with no sums over intermediate spectator momenta actually appearing.
The same comment applies below for all instances of this Kronecker delta.}
to an overall $\delta_{k p}$.

The non-trivial sum is that contained in $F$.
Naively one might say that, since $F$ contains a sum-integral
difference [see Eq.~(\ref{eq:Fdef3})], $F\to 0$ when $L\to \infty$.
After all, $F$ is the source of finite-volume corrections.
However, as already noted, this conclusion depends on the
pole prescription chosen to define the sum before the limit is evaluated. 
If we choose the $\PV$ prescription for the sum,
then, indeed, $F$ vanishes in the limit:\footnote{%
This is because the $\rho$ term in Eq.~(\ref{eq:Fdef2})
is exactly what is required to turn the integral with
the $i\epsilon$ prescription in Eq.~(\ref{eq:Fdef3}) into
an integral with the $\PV$ prescription. See Ref.~\cite{us} for
further discussion.}
\begin{equation}
\lim_{L \rightarrow \infty}\bigg \vert_{\PV} F  = 0 \,,
\end{equation}
Here we denote the choice of pole prescription by the subscript.
On the other hand, if we define the limit by regulating the sum with an 
$i\epsilon$ prescription, then
it is $F^{i\epsilon}$ of Eq.~(\ref{eq:Fdef3}) that vanishes, so that
\begin{equation}
\label{eq:Flimit}
\lim_{L \rightarrow \infty} \bigg \vert_{i \epsilon} F  
= \delta_{k'k}\; \rho_{\ell' m';\ell m}(\vec k) \equiv \rho \,,
\end{equation}
where $\rho_{\ell' m;\ell m}(\vec k)$ is the phase space
matrix given in Eq.~(\ref{eq:rhodef}). We stress that, in the
following, $\rho$ without an argument is a matrix with full $k\ell m$ 
indices while $\rho(\vec k)$ has only $\ell m$ indices.

Using these results we can immediately determine the infinite-volume
limit of $\ML$. If we use the $\PV$ prescription for the limit of sums,
then 
\begin{equation}
\lim_{L \rightarrow \infty} \bigg \vert_{\PV} \ML  
=\K \,.
\label{eq:PVlimit}
\end{equation}
Above threshold, $\PV$ is equivalent to the principal value prescription.
Thus Eq.~(\ref{eq:PVlimit}) reproduces the well known result that 
evaluating scattering diagrams using the
PV prescription in loops (i.e. dropping the imaginary part) leads to the
K-matrix. This is not, however, the infinite-volume limit that we desire.
If we instead use the $i\epsilon$ prescription we obtain
\begin{equation}
\lim_{L \rightarrow \infty} \bigg \vert_{i \epsilon} \ML = \M
=\frac{1}{1+\K \rho}\K\,.
\label{eq:infvolML}
\end{equation}
Above threshold, this is indeed the standard relation between $\M$ and
$\K$, with $\rho$ adding in the cuts needed for unitarity.
This provides a check that we are taking the infinite-volume limit  correctly.
Below threshold (i.e. with the spectator momentum chosen so that the
scattering pair are below threshold), this result provides an
alternative expression for $\K$  in terms of $\M$ analytically continued below threshold:
 \begin{equation}
 \K^{-1} = \M^{-1}-\rho\,.
 \end{equation}
Note that, although $\K$ and
$\rho$ depend on the cut-off function $H(\vec k)$, this dependence
cancels in $\M$.

\bigskip
We now apply the same limiting procedure to our result for $\MthL$, 
Eq.~(\ref{eq:MthLres}). To do so we need some new notation. Since $\vec p$ and $\vec k$ become continuous
variables, while angular-momentum indices remain unchanged, we adopt
a hybrid notation for $\Kdf$ and ${\cal D}^{(u,u)}$, e.g.
\begin{equation}
\label{eq:Kdfrelab}
{\cal K}_{3,{\rm df};p \ell' m';k \ell m}
=
{\cal K}_{3,{\rm df};\ell' m';\ell m}(\vec p, \vec k)
\,.
\end{equation}
We stress here that ${\cal K}_{3,{\rm df}}$ was already defined for 
arbitrary $\vec p$ and $\vec k$ above so that Eq.~(\ref{eq:Kdfrelab}) 
is only a relabeling. 
Similarly, we make the continuous spectator momentum 
an argument of $\M$,
and drop the Kronecker delta from its definition:
\begin{equation}
\lim_{L \rightarrow \infty} \bigg \vert_{i \epsilon} 
{\cal M}_{2,L;p \ell' m';k \ell m}
=
\delta_{p k} {\cal M}_{2;\ell' m';\ell m}(\vec p \,)
\,.
\end{equation}
This is similar to the notation for $\rho$ in Eq.~(\ref{eq:Flimit}).
Finally, for $\Mth$ and its divergent part $\mathcal D$, which, after symmetrization, are functions
of  angular variables rather than spherical harmonics
[as discussed in Sec.~\ref{sec:B2} around Eq.~(\ref{eq:symmMthL})], we use
\begin{equation}
\lim_{L \rightarrow \infty} \bigg \vert_{i \epsilon} 
\MthL(\vec p, \hat a'^*;\vec k,\hat a^*)
\equiv
\Mth(\vec p,\hat a'^*;\vec k,\hat a^*)
\,,
\end{equation}
and similarly for $\mathcal D$.

Our task is now to evaluate the limit in
\begin{multline}
i{\cal M}_{3}(\vec p,\hat a'^*;\vec k,\hat a^*) =
\lim_{L \rightarrow \infty} \bigg \vert_{i \epsilon} 
{\cal S}\bigg\{ i{\cal D}_L^{(u,u)}
\\
+
\left[\frac13 + \frac{1}{1\! -\! i \ML i G} i \ML i F\right] 
i \Kdf \frac{1}{1\!-\! i F_3 i \Kdf} 
\left[\frac13 + \frac{i F}{2 \omega L^3}\frac{1}{1 \!-\! i \ML i G} 
i \ML (2 \omega L^3) \right ] \bigg \} \,.
\end{multline}
We can interchange the symmetrization and the taking of 
$L\to\infty$ because the changes of 
variables involved in symmetrization are independent of $L$.
We thus begin by determining the $L\to\infty$ limit of 
$\DL^{(u,u)}$, given in Eq.~(\ref{eq:DLuudef}). To do so we replace $\ML$ with $\M$,
and introduce an infinite-volume version of $G$
\begin{equation}
\label{eq:Ginfdef}
G_{\ell' m' ; \ell m}^\infty(\vec p, \vec k)
 \equiv
\left(\frac{k^*}{q_p^*}\right)^{\ell'} 
\frac{4 \pi Y_{\ell' m'}(\hat  k^*) 
H(\vec p\,) H(\vec k\,) Y_{\ell m}^*(\hat p^*)} 
{2 \omega_{kp} (E - \omega_k - \omega_p - \omega_{kp}+i\epsilon)}
\left(\frac{p^*}{q_k^*}\right)^\ell \,.
\end{equation}
This differs from the matrix form of $G$,
given in Eq.~(\ref{eq:Gdef}),
simply by the removal of a factor of $1/({2 \omega_k L^3})$\,.
This factor either cancels the explicit
$(2 \omega L^3)$ in $\DL^{(u,u)}$, or converts sums over
internal spectator momentum indices into integrals.
For example the first two terms in the geometric series
obtained by expanding $\DL^{(u,u)}$ have the following limits:
\begin{align}
\lim_{L \rightarrow \infty} \bigg \vert_{i \epsilon} 
i \M iG i\M (2\omega L^3)
&=
i \M(\vec p) i G^\infty(\vec p,\vec k) i\M(\vec k) \,,
\\
\lim_{L \rightarrow \infty} \bigg \vert_{i \epsilon} 
i \M iG i\M iG i\M (2\omega L^3)
&=
\int_s \frac1{2\omega_s} i \M(\vec p) i G^\infty(\vec p,\vec s) i\M(\vec s) 
i G^\infty(\vec s,\vec k) i\M(\vec k)\,,
\end{align}
where angular momentum indices are implicit.
Here we are using $\int_s \equiv \int d^3s/(2\pi)^3$.
Repeating for higher-order terms we find that the
infinite-volume form of $\DL^{(u,u)}$,
\begin{equation}
\lim_{L \rightarrow \infty} \bigg \vert_{i \epsilon}
{\cal D}^{(u,u)}_{L;p \ell' m';k \ell m}
\equiv
{\cal D}^{(u,u)}_{\ell' m; \ell m}(\vec p, \vec k)\,,
\end{equation}
satisfies the integral equation
\begin{equation}
\label{eq:Duuinteq}
i \mathcal D^{(u,u)}(\vec p, \vec k) =
i \M(\vec p) i G^\infty(\vec p,\vec k) i\M(\vec k) 
+
\int_s \frac1{2\omega_s} i \M(\vec p) i G^\infty(\vec p,\vec s \,) 
i{\cal D}^{(u,u)}(\vec s,\vec k) \,.
\end{equation}
Symmetrizing following Eq.~(\ref{eq:Sdef}) leads to 
\begin{equation}
{\cal D}(\vec p,\hat a'^*;\vec k,\hat a^*) = {\cal S}\left\{ {\cal D}^{(u,u)}(\vec p,\vec k)
\right\}\,.
\end{equation}
This quantity contains all the physical divergences in $\Mth$,
i.e.~divergences that occur above threshold for physical external momenta.
The difference
\begin{equation}
{\cal M}_{{\rm df},3}(\vec p,\hat a'^*;\vec k,\hat a^*) 
=
\Mth(\vec p,\hat a'^*;\vec k,\hat a^*) - {\cal D}(\vec p,\hat a'^*;\vec k,\hat a^*)
\label{eq:Mdfthdef}
\end{equation}
is free of such divergences, and can, if desired, be expanded in spherical harmonics.

To determine the $L\to\infty$ limit of the remainder of $\MthL$ we need
to evaluate
\begin{equation}
\lim_{L \rightarrow \infty} \bigg \vert_{i \epsilon}  
\bigg\{
 \left[\frac13 + \frac{1}{1 - i \ML i G} i \ML  i F \right ] 
i \Kdf \frac{1}{1 - i F_3  i \Kdf} 
\left[\frac13 + \frac{i F}{2 \omega L^3}
\frac{1}{1 - i \ML i G} i \ML (2 \omega L^3) \right ] \bigg \}\,,
\label{eq:lastjob}
\end{equation}
where $F_3$ is defined in Eq.~(\ref{eq:FthdefML}).
We first take the infinite-volume limit of the central part of
(\ref{eq:lastjob}), i.e.
\begin{equation}
\lim_{L \rightarrow \infty} \bigg \vert_{i \epsilon} 
\left[i \Kdf \frac{1}{1- i F_3 i \Kdf}\right]_{p \ell' m;k \ell m} 
\equiv  
i \mathcal T_{\ell' m';\ell m}(\vec p, \vec k) 
\,.
\end{equation}
Rewriting $F_3$ as
\begin{equation}
i F_3 = \frac{iF}{2\omega L^3}\left[
\frac13 + i \ML iF + i \DL^{(u,u)} \frac{iF}{2\omega L^3}\right]
\,,
\end{equation}
we find that ${\cal T}$ is given by the solution to the integral equation
\begin{equation}
i \mathcal T(\vec p, \vec k) 
=
i \Kdf(\vec p, \vec k)  + 
\int_s  \int_r 
i \Kdf(\vec p, \vec s\,) \frac{i \rho(\vec s\,)}{2 \omega_s}
 {\cal L}^{(u,u)}(\vec s, \vec r\,) i \mathcal T(\vec r, \vec k)  \,,
\label{eq:Tres}
\end{equation}
where 
\begin{equation}
 \mathcal L^{(u,u)}(\vec p, \vec k) =  
\left(\frac13 + i \M(\vec p\,) i \rho(\vec p\,) \right)
(2 \pi)^3 \delta^3(\vec p - \vec k) 
+ i \mathcal D^{(u,u)}(\vec p, \vec k) \frac{i \rho(\vec k)}{2 \omega_k} \,,
\label{eq:Luures}
\end{equation}
and one must additionally
enforce that ${\cal T}$ is symmetric under particle interchange symmetry
on its left-hand argument.\footnote{%
One way to do this would be to symmetrize the matrix appearing between
$\Kdf$ and ${\cal T}$.}
The factors of $\rho(\vec k)$ arise from the infinite-volume limit of $F$.\footnote{%
One would expect the final result to involve integrals over Lorentz-invariant
three-body phase space, and indeed this is the case.
The implicit sum over angular momentum indices can be rewritten 
as an angular integral in the scattering pair's CM frame.
This integral, combined with that over the spectator momentum,
gives three-body phase space (extended to include subthreshold particles) as long as one includes
the Jacobian factor $\rho(\vec k)/2\omega_k$.}

What remains to be considered are
the combinations at the left- and right-hand ends of
Eq.~(\ref{eq:lastjob}). These become integral operators in the
infinite-volume limit. Putting everything together we find
\begin{equation}
i{\cal M}_{3}(\vec p, \hat a'^*; \vec k,\hat a^*)
= i{\cal D}(\vec p,\hat a'^*;\vec k,\hat a^*) +
{\cal S}
\left\{
\int_s \int_r 
 {\cal L}^{(u,u)}(\vec p,\vec s\,)
i {\cal T}(\vec s,\vec r\,)
 {\cal R}^{(u,u)}(\vec r,\vec k\,)\right\}
\,,
\label{eq:final}
\end{equation}
where ${\cal R}^{(u,u)}$ is the reflection of ${\cal L}^{(u,u)}$:
\begin{equation}
 \mathcal R^{(u,u)}(\vec p, \vec k) =  
\left(\frac13 +  i \rho(\vec p\,) i \M(\vec p\,) \right)
(2 \pi)^3  \delta^3(\vec p - \vec k) 
+  \frac{i \rho(\vec p \,)}{2 \omega_p}  
i \mathcal D^{(u,u)}(\vec p, \vec k\,) \,.
\label{eq:Ruures}
\end{equation} 

Eq.~(\ref{eq:final}) is our final result. 
It shows that $\Mth$ can be obtained
from $\Kdf$ by solving two integral equations
(first for ${\cal D}^{(u,u)}$, which requires $\M$, and
then for ${\cal T}$), applying two integral
operators to ${\cal T}$, symmetrizing,
and then adding in the divergent part ${\cal D}$.
We note that all the integrals appearing above have a finite
range because all integrands include the cutoff function $H$.
This may be helpful when making numerical approximations.

\section{Expressing $\Kdf$ in terms of $\Mth$}
\label{sec:KfromM}

In this brief section we show how the result (\ref{eq:final}) can
be inverted so as to obtain $\Kdf$ given $\Mth$ (and $\M$).
This will be useful if one has a model for $\Mth$ and wants to determine
the prediction for $\Kdf$, which can then be inserted into the
quantization condition of Ref.~\cite{us} in order to predict the
finite-volume spectrum.

We begin by rewriting Eq.~(\ref{eq:final}) in terms of symmetric quantities:
\begin{multline}
i{\cal M}_{{\rm df},3}(\vec p, \hat a'^*;\vec k,\hat a^*)
= 
\int_s  \int_{\hat b'^*} \int_r \int_{\hat b^*}
\left\{
(2\pi)^3 \delta^3(\vec p - \vec s\,)4\pi\delta^2(\hat a'^*-\hat b'^*)
+ \Delta_{\cal L}(\vec p,\hat a'^*;\vec s,\hat b'^*) \right\}
\\ \times
i {\cal T}(\vec s,\hat b'^*;\vec r,\hat b^*)
\, \left\{
(2\pi)^3 \delta^3(\vec r -\vec k\,)4\pi\delta^2(\hat b^*-\hat a^*)
+ \Delta_{\cal R}(\vec r,\hat b^*;\vec k,\hat a^*) \right\}
\,.
\label{eq:final2}
\end{multline}
Here the angular averages and corresponding delta-functions are defined by
\begin{equation}
\int_{\hat b} \equiv \frac1{4\pi} \int d \Omega_{\hat b}\,,\  \  {\rm and}\  \
\int_{\hat b} 4 \pi \delta^2(\hat b - \hat a) f(\hat b) \equiv f(\hat a)\,.
\end{equation}
These angular averages arise from reexpressing the sums over repeated $\ell m$ indices
as angular integrals.
We have also changed from spherical harmonic indices  to explicit angular dependence
for the quantity ${\cal T}$, using our standard normalization [see, e.g., Eq.~(\ref{eq:MthLangtosph})].
Finally, the quantity $\Delta_{\cal R}$  is defined by 
\begin{align}
(2\pi)^3 \delta^3(\vec p-\vec k) 4\pi \delta^2(\hat a'^*-\hat a^*)
+ {\Delta}_{\mathcal R}(\vec p,\hat a'^*;\vec k,\hat a^*) 
&\equiv  \frac13  {\cal S}\left\{ 
{\cal R}^{(u,u)}(\vec p,\vec k\,) \right\}
\,,
\end{align}
with $\Delta_{\cal L}$ defined analogously.
Note that since ${\cal R}^{(u,u)}$ and ${\cal L}^{(u,u)}$ are singular quantities (due to the
delta functions and the poles in ${\cal D}^{(u,u)}$) they must be symmetrized in the
manner described by Eq.~(\ref{eq:Sdef}). This is why we must change to angular variables.

To obtain Eq.~(\ref{eq:final2}) from Eq.~(\ref{eq:final}) we have, apart from the change to
angular variables, used the fact that ${\cal T}$ is symmetric under independent particle
exchange on both its initial and final arguments. This allows us to symmetrize
${\cal L}^{(u,u)}$ and ${\cal R}^{(u,u)}$ on both the outer and inner arguments,
which is what is done by the operation of ${\cal S}$. To avoid over counting we must
include the factor of $1/3$.
We note that, although $\Delta_{\cal L}$ and $\Delta_{\cal R}$
are complicated quantities, one can in principle construct them
given $\M$.

We now invert Eq.~(\ref{eq:final2}) in stages. The first
step is to construct ${\cal M}_{{\rm df},3}$ from $\Mth$
using Eq.~(\ref{eq:Mdfthdef}). Next we invert the factors on the
left and right sides of ${\cal T}$ in (\ref{eq:final2}).
This is done using kernels $I_{\cal L}$ and
$I_{\cal R}$ that solve the integral equations
\begin{align}
\label{eq:ILdef}
 I_{\cal L}(\vec p,\hat a'^*;\vec k,\hat a^*) &= (2\pi)^3 \delta(\vec p-\vec k\,)4\pi\delta^2(\hat a'^*-\hat a^*)
-  \int_s \int_{\hat b^*}
I_{\cal L}(\vec p,\hat a'^*;\vec s,\hat b^*) \Delta_{\cal L}(\vec s,\hat b^*;\vec k,\hat a^*) \,,
\\
\label{eq:IRdef}
 I_{\cal R}(\vec p,\hat a'^*;\vec k,\hat a^*) &= (2\pi)^3 \delta(\vec p-\vec k\,)4\pi\delta^2(\hat a'^*-\hat a^*)
-  \int_s \int_{\hat b^*}
\Delta_{\cal R}(\vec p,\hat a'^*;\vec s,\hat b^*) I_{\cal R}(\vec r,\hat b^*;\vec k,\hat a^*) 
\,.
\end{align}
One then finds that
\begin{equation}
4\pi Y^*_{\ell' m'}(\hat a'^*)i{\cal T}(\vec p,\vec k\,)_{\ell' m',\ell m} Y_{\ell m}(\hat a^*)
= \int_s \int_{\hat b'^*} \int_r\int_{\hat b^*}
I_{\cal L}(\vec p,\hat a'^*;\vec s,\hat b'^*) i {\cal M}_{{\rm df},3}(\vec s,\hat b'^*;\vec r,\hat b^*)
I_{\cal R}(\vec r,\hat b^*;\vec k,\hat a^*)
\,,
\end{equation}
where we have used the fact that ${\cal T}$ is non-singular to keep it expressed with
spherical harmonic indices.

Finally, we reconstruct $\Kdf$ from ${\cal T}$ by rewriting Eq.~(\ref{eq:Tres}) as
\begin{equation}
i \Kdf(\vec p, \vec k)   
=
i \mathcal T(\vec p, \vec k) -
\int_s  \int_r 
i \Kdf(\vec p, \vec s\,) \frac{i \rho(\vec s\,)}{2 \omega_s}
 {\cal L}^{(u,u)}(\vec s, \vec r\,) i \mathcal T(\vec r, \vec k)  \,,
\label{eq:Tres2}
\end{equation}
and then solving for $\Kdf$. This must be done
with the added condition that $\Kdf$ is symmetric under
particle interchange symmetry.

\section{Simplifying Cases}
\label{sec:simp}

In this section we describe two approximations in which the
relation between $\Mth$ and $\Kdf$ simplifies substantially.
Such simplifications are likely to be the first step towards using
this formalism in practice.
Both approximations were discussed in Ref.~\cite{us}.

We begin by considering the result in the approximation that only the $s$-wave components 
of $\M$ and $\Kdf$ are nonzero---which we call the s-wave approximation. 
In this case the quantities $\Kdf$, ${\cal D}^{(u,u)}$,
$\mathcal T$, $\mathcal L$, $\mathcal R$ and $\M$ all are non-zero only
for $\ell=m=0$, so the spherical harmonic indices can be dropped.
This implies that Eqs.~(\ref{eq:Duuinteq}), (\ref{eq:Tres}), (\ref{eq:Luures})
and (\ref{eq:Ruures}) retain their forms, 
except that now all quantities are no longer matrices.
In addition,  $G^\infty$ becomes the function
\begin{equation}
G^\infty(\vec p, \vec k\,)
= \frac{H(\vec p) H(\vec k)}{2\omega_{kp}(E-\omega_k-\omega_p-\omega_{kp}+i\epsilon)}
\,.
\end{equation}
The final equation (\ref{eq:final}) is replaced by
\begin{equation}
i \mathcal M_3(\vec p, \hat a'^*; \vec k, \hat a^*) =
 \mathcal S \bigg \{ 
i \mathcal D^{(u,u)}_{3}(\vec p, \vec k) + \int_s \int_r
\mathcal L^{(u,u)}(\vec p, \vec s\,) i \mathcal T(\vec s, \vec r \,) \mathcal R^{(u,u)}(\vec r, \vec k\,) 
   \bigg \} \,,
\end{equation}
where symmetrization is defined as before [Eq.~(\ref{eq:Sdef})] except that now the quantity
being symmetrized has no dependence on $\hat a'^*$ and $\hat a^*$. Nevertheless, such
dependence is introduced by the symmetrization, and thus $\Mth$ maintains its
dependence on the full set of kinematic variables (albeit in a simplified form).

Next we consider the more restrictive approximation---referred to in Ref.~\cite{us} as
the isotropic approximation---in which the s-wave approximation
is extended by assuming that $\Kdf(\vec p, \vec k)$ 
depends only on the total three-particle CM energy $E^*$, 
\begin{equation}
\Kdf(\vec p, \vec k) \longrightarrow \Kdf^{\rm iso}(E^*)\,.
\end{equation}
In this case, it is shown in Ref.~\cite{us} that $\Kdf^{\rm iso}$ can be obtained
from the spectrum using a simple algebraic equation [Eq.~(38) of \cite{us}].
The integral equations which determine $\Mth$ are all at fixed $E$ and $\vec P$ and thus
at fixed $E^*$. Thus $\Kdf^{\rm iso}$ is simply a constant, which simplifies some of the results.

We find that ${\cal D}^{(u,u)}$ is the same as in the s-wave approximation,
while ${\cal L}$ and ${\cal R}$ simplify to
\begin{align}
 \mathcal L(\vec p) & = 
  \left[\left(\frac13 + i \M(\vec p\,) i \rho(\vec p\,) \right)
  + \int_s i \mathcal D^{(u,u)}(\vec p, \vec s)  \frac{i \rho(\vec s)}{2\omega_s} \right ] \,, 
  \\
 \mathcal R( \vec k) & =   
 \left[\left(\frac13 +  i \rho(\vec k\,) i \M(\vec k\,) \right) 
 + \int_r \frac{i \rho(\vec r \,)}{2\omega_r}  i \mathcal D^{(u,u)}(\vec r, \vec k) \right ]\,.
\end{align}
Introducing the new quantity
\begin{align}
i  F_3^\infty  & =  \int_s \frac{i\rho(\vec s\,)}{2\omega_s}\left[ \frac13+ i \M(\vec s\,) i \rho(\vec s\,) \right] 
+ \int _s\int_r \Big[ \frac{i \rho(\vec s\,)}{2\omega_s} i \mathcal D^{(u,u)}(\vec s, \vec r\,) 
\frac{i \rho(\vec r)}{2\omega_r}\Big ]  \,,
\end{align}
the final result is
\begin{equation}
i \mathcal M_3(\vec p, \hat a'^*; \vec k, \hat a^*) = \mathcal S \bigg \{
 i \mathcal D^{(u,u)}_{3}(\vec p, \vec k)
 +   \mathcal L(\vec p)   i \Kdf^{\rm iso} \frac{1}{1 - i F_3^\infty i \Kdf^{\rm iso}}  \mathcal R(\vec k) 
 \bigg\} \,.
\end{equation}
Thus, in this approximation, the only integral equation that needs to be solved is
that for ${\cal D}^{(u,u)}$. Apart from that, the result for $\Mth$ is obtained simply by
doing integrals, all of which are over a finite range due to the cutoff function $H$.

\section{Conclusions and outlook}
\label{sec:conc}

 In this work we have derived a relation between the non-standard three-particle K-matrix,
 $\mathcal K_{{\rm df}, 3}$, and the standard three-to-three scattering amplitude, $\mathcal M_3$. 
 This completes the formalism relating the finite-volume three-particle spectrum 
 to scattering observables and removes the central drawback of the formalism presented in Ref.~\cite{us}.

Future work will be dedicated to providing non-trivial checks of the formalism completed here, as well as 
assessing its utility. Concerning the former aim, we are currently completing a study of the three-particle 
quantization condition for $\vec P=0$ and for energy close to three-particle threshold, $E \approx 3m$. 
Here we can compare our result to those obtained using 
non-relativistic quantum mechanics~\cite{Beane:2007qr,Tan:2007},
and using perturbation theory in a $\lambda \phi^4$ theory~\cite{ourpt}.
As we describe in our upcoming work, we find nontrivial agreement between our general
result in this limit and the results of the alternative analyses. 

Concerning the utility of this framework, one important focus will be on solving the 
integral equations relating $\mathcal K_{{\rm df},3}$ and $\mathcal M_3$. 
A possible direction that one might pursue follows from the observation that, 
with total energy and momentum fixed, the phase space for any number of particles is compact. 
This means that smooth functions which depend on this phase space, such as $\mathcal M_2$ in the case of two-particles and $\mathcal K_{{\rm df}, 3}$ in the case of three, 
must be expressible as an infinite series of generalized harmonics. 
In the two-particle case this is the standard decomposition in spherical harmonics, leading to the parametrization of $\mathcal M_2$ in terms of an energy-dependent phase shift for each partial wave. Identifying an analogous decomposition for $\mathcal K_{{\rm df}, 3}$ would be a first step
in providing a well-motivated parametrization to extract this quantity from the spectrum and 
would likely also simplify the procedure for constructing $\mathcal M_3$.

Finally, we stress that the most important function of this work is as a stepping stone to the formalism required to
describe more complicated and more physically interesting systems. We thus aim, in future work, to remove the
restrictions listed in the introduction and provide a formal result for all possible three-hadron states. 
For example, to accommodate nucleon channels we must incorporate non-zero intrinsic spin into this analysis. 
This has been studied thoroughly in the two-particle sector in Refs.~\cite{Li:2012bi,Briceno:2014oea, Briceno:2015csa}. We expect the methods developed there will facilitate the generalization to three-particles. 
A further necessary extension is to accommodate two-to-three transitions, 
as well as transitions involving multiple two- and three-particle channels, 
possibly containing non-identical and non-degenerate particles. 
Although this will introduce many complicating details to the analysis, we expected that much of the technology developed here can be reapplied. 
Finally, we hope to extend this work to apply to systems with two-body bound states and resonances. 
This will require accommodating new finite-volume effects arising from bound-state and resonance poles.

\section*{Acknowledgments}

We thank Ra\'ul Brice\~no and Akaki Rusetsky for discussions. SRS was supported in part
by the United States Department of Energy grants DE-FG02-96ER40956 and DE-SC0011637. 

\appendix

\section{Definitions}
\label{app:defs}

We collect here definitions from Ref.~\cite{us} that are not given in the main text.
The matrices appearing in $F_3$, Eq.~(\ref{eq:FthdefML}) are
\begin{align}
\label{eq:omegamatdef}
\left[ {2 \omega L^3} \right]_{k'\ell'm';k\ell m} & \equiv
\delta_{k'k} \delta_{\ell'\ell} \delta_{m'm} {2 \omega_k L^3} 
\,,\\ 
\label{eq:Gdef}
G_{p \ell' m' ; k \ell m} 
& \equiv
\left(\frac{k^*}{q_p^*}\right)^{\ell'} 
\frac{4 \pi Y_{\ell' m'}(\hat  k^*) 
H(\vec p\,) H(\vec k\,) Y_{\ell m}^*(\hat p^*)} 
{2 \omega_{kp} (E - \omega_k - \omega_p - \omega_{kp})}
\left(\frac{p^*}{q_k^*}\right)^\ell 
\frac{1}{2 \omega_k L^3} \,, 
\\
\label{eq:Fdef1}
F_{k' \ell'm';k\ell m} 
& \equiv 
\delta_{k'k} F_{\ell'm';\ell m}(\vec k)\,,
\\
\label{eq:Fdef2}
F_{\ell' m';\ell m}(\vec k)
&=
F^{i\epsilon}_{\ell' m';\ell m}(\vec k)
+
\rho_{\ell'  m'; \ell m}(\vec k) \,,
\\
\label{eq:Fdef3}
F^{i\epsilon}_{\ell' m';\ell m}(\vec k)
&=
\frac12 \left[\frac{1}{L^3} \sum_{\vec p} - \int \frac{d^3p}{(2\pi)^3} \right] 
\frac{{4 \pi} Y_{\ell' m'}(\hat p^*) Y_{\ell m}^*(\hat p^*) H(\vec k)
  H(\vec p\,)H(\vec b_{kp})}
{2 \omega_p 2 \omega_{kp}(E - \omega_k - \omega_p -  \omega_{kp} + i \epsilon)}
\left(\frac{p^*}{q_k^*}\right)^{\ell+\ell'} \,.
\end{align}
where the sum over $\vec p$ in $F$ runs over all finite-volume momenta. 
Here $\vec b_{kp}=\vec P-\vec k-\vec p$ is the momentum of the third particle,
and $\omega_{kp}=\sqrt{\vec b_{kp}^2+m^2}$ is its on shell energy.
The momentum $\vec p^{\;*}$ is the result of boosting $\vec p$ to the frame
in which the non-spectator pair 
[which has four-momentum $P_2$, see Eq.~(\ref{eq:P2def})]
is at rest. If all three particles are on shell, then 
$|\vec p^{\;*}|\equiv p^*=q_k^*$, 
where $q_k^*$ is given in Eq.~(\ref{eq:qkstardef}).
The quantities $\vec k^*$ and $q_p^*$ are obtained similarly by
interchanging the roles of $\vec k$ and $\vec p$.
The matrix $\rho$ is a phase-space factor for the non-spectator pair,
defined by
\begin{align}
\label{eq:rhodef}
\rho_{\ell' m';\ell m}(\vec k)& \equiv 
\delta_{\ell' \ell} \delta_{m' m} H(\vec k) \tilde\rho(P_2)\,,
\\
\tilde\rho(P_2) &\equiv \frac{1}{16 \pi  \sqrt{P_2^2}} \times
\begin{cases} 
-  i \sqrt{P_2^2/4-m^2} & (2m)^2< P_2^2 \,, 
\\ 
\vert \sqrt{P_2^2/4-m^2} \vert &   0< P_2^2 \leq (2m)^2 \,.
\end{cases}
\label{eq:rhodef2}
\end{align}

Here, and in the definitions of $G$ and $F$,
$H$ provides a smooth ultraviolet cut-off. We require
\begin{equation}
\label{eq:HvalA}
 H(\vec k) = 
\begin{cases}
0 \,, & P_2^{2} \le 0 \,;
\\ 
1 \,, & (2m)^2 < P_2^{2}\,,
\end{cases}
\end{equation}
where the first condition removes unphysical boosts
and the second ensures that the cut-off does not change the
contributions from on-shell intermediate states.
In the intermediate region, \(0 < P_2^{2} < (2m)^2\), 
$H(\vec k)$ interpolates smoothly between 0 and 1. An example of a function which satisfies all requirements is
\begin{equation}
\label{eq:Hdef}
H(\vec k) \equiv J(P_2^{2}/[4m^2]) \,,
\end{equation}
with
\begin{equation}
\label{eq:Jdef}
J(x) \equiv
\begin{cases}
0 \,, & x \le 0 \,; 
\\ 
\exp \left( - \frac{1}{x} \exp \left [-\frac{1}{1-x} \right] \right ) \,, 
& 0<x \le 1 \,; 
\\ 
1 \,, & 1<x \,.
\end{cases}
\end{equation}

The definitions in Eqs.~(\ref{eq:omegamatdef})-(\ref{eq:rhodef2})
are used in Ref.~\cite{us} solely with the spectator momenta ($\vec k$, $\vec k'$ and $\vec p$)
as well as the total momentum ($\vec P$) all being finite-volume momenta.
As noted repeatedly in the main text, in this work we need to consider the extension to the case that
these momenta are not in the finite-volume set. This presents no problems for any of the above
definitions---all extend smoothly to general values of $\vec k$, $\vec k'$, $\vec p$ and $\vec P$.

\bibliographystyle{apsrev4-1} 
\bibliography{ref} 

\end{document}